\documentclass[11pt]{article}

\usepackage{cite}
\usepackage[dvipsnames]{xcolor}
\usepackage{empheq}
\usepackage{amsmath}
\usepackage{bbold}
\usepackage{amssymb}
\usepackage{ulem,xpatch}
\usepackage{titling}
\usepackage{geometry} 
\usepackage{tabularx}
\usepackage{physics}
\usepackage{siunitx}
\newcommand{\degree}{\ensuremath{^\circ}}

\newcommand{\hide}[1]{\relax}

\newcommand{\nocontentsline}[3]{}
\newcommand{\tocless}[2]{\bgroup\let\addcontentsline=\nocontentsline#1{#2}\egroup}

\title{A soft-clamped topological waveguide for phonons}
\author{
    Xiang Xi$^{1,2}$,
    Ilia Chernobrovkin$^{1,2}$,
    Jan Ko\v{s}ata$^{3,\ast}$,
    Mads B. Kristensen$^{1,2}$,\\
    Eric Langman$^{1,2}$,
    Anders S. S{\o}rensen$^{1,2}$,
    Oded Zilberberg$^{4}$,
    Albert Schliesser$^{1,2,\dagger}$
}

\begin{document}
\maketitle

\begin{enumerate}
 \item Niels Bohr Institute, University of Copenhagen, Copenhagen, Denmark
 \item Center for Hybrid Quantum Networks (Hy-Q), Niels Bohr Institute, University of Copenhagen, Copenhagen, Denmark.
 \item Institute for Theoretical Physics, ETH Z\"urich, Z\"urich, Switzerland
 \item Department of Physics, University of Konstanz, Konstanz, Germany\\
\end{enumerate}
\normalsize{$^\ast$ current address: Hitachi Energy Research, Baden-D\"attwil, Switzerland}\\
\normalsize{$^\dagger$ email:albert.schliesser@nbi.ku.dk}

\begin{abstract}
Topological insulators were originally discovered for electron waves in condensed matter systems.
Recently this concept has been transferred to bosonic systems such as photons\cite{Ozawa2019} and phonons\cite{Shah2024}, which propagate in materials patterned with artificial lattices that emulate spin-Hall physics.
This work has been motivated, in part, by the prospect of topologically protected transport along edge channels in on-chip circuits\cite{Hafezi2011, Shah2024}.
Importantly, even in principle, topology protects propagation against backscattering, but not against loss, which has remained limited to the dB/cm-level for phonon waveguides, be they topological\cite{Cha2018Nature, yu2020topo, Xi2021, Hatanaka2024} or {not\cite{Hatanaka2014, Hatanaka2015, Cha2018, fu2019phononic, Patel2018, Romero2019, Mayor2021, xu2022, Feng2023, Chen2023_om, Zhang2024, Bicer2024}}.
Here, we combine advanced dissipation engineering\cite{Sementilli2022,Engelsen2024}, in particular the recently introduced method of soft-clamping\cite{Tsaturyan2017}, with the concept of a valley-Hall topological insulator for {phonons\cite{Lu2017,yan2018chip,jwma2021,Zhang2022,Ren2022topo}}.
This enables on-chip phononic waveguides with propagation losses of 3~dB/km at room temperature, orders of magnitude below any previous chip-scale devices.
For the first time, the low losses also allow us to accurately quantify backscattering protection in a topological phonon waveguide, using high-resolution ultrasound spectroscopy. 
We infer that phonons follow a sharp, $120^{\degree}$-bend with a $99{.}99\%$-probability instead of being scattered back, and less than one phonon in a million is lost.
The extraordinary combination of features of this novel platform suggest applications in classical and quantum signal routing, processing, and storage.

\end{abstract}

\clearpage
\newpage

%\section{Introduction}

\begin{figure}
\centering
\begin{center}
\includegraphics[width=0.8\linewidth]{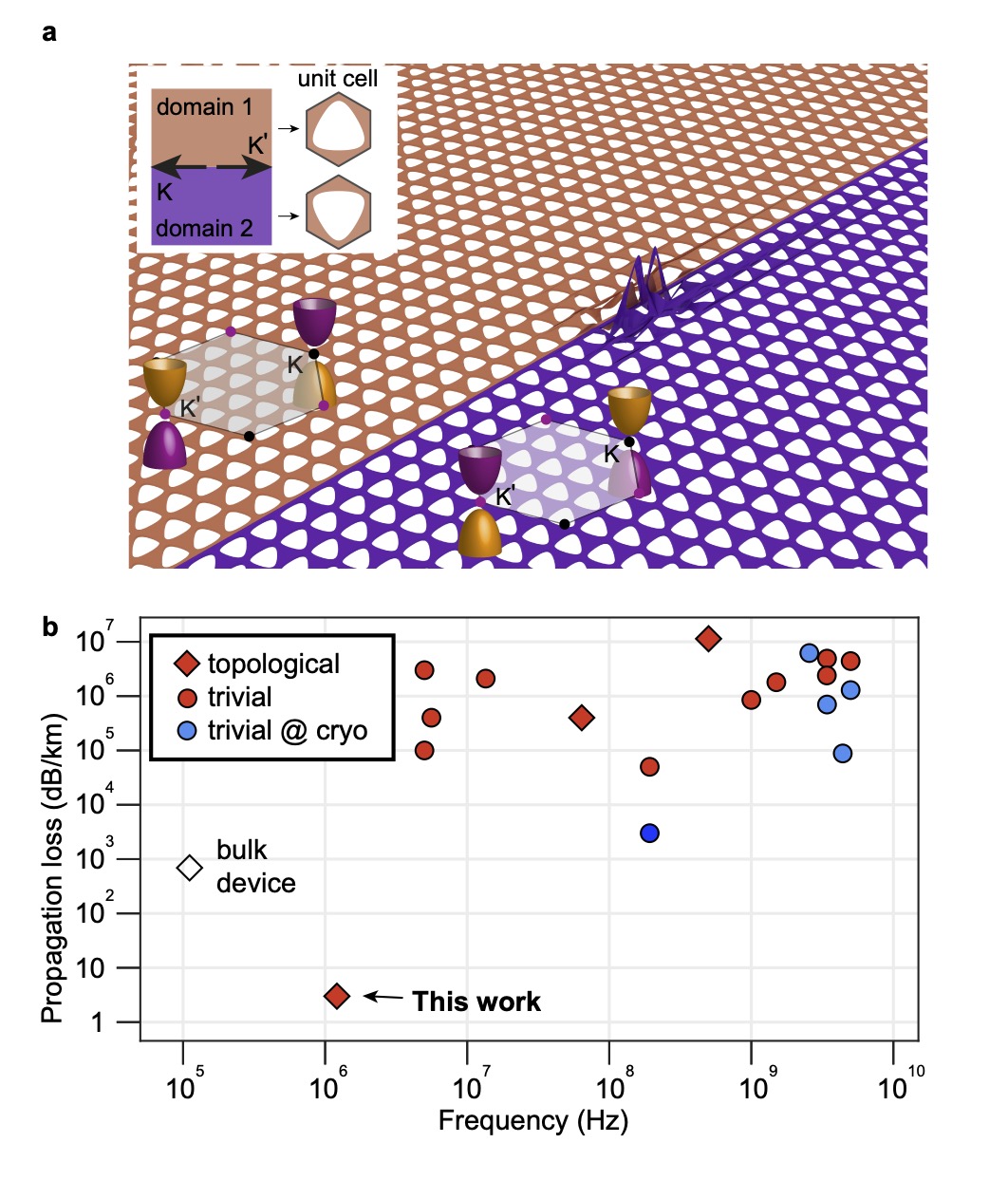}

\caption{{\bf A topological waveguide for phonons with ultralow loss.}
(a) Rendering of an out-of-plane vibration mode guided along the interface of two topologically distinct domains, realised by patterning a thin, stressed membrane with a regular array of holes.
Upper inset shows the two different phononic crystal unit cells of the two domains. 
Lower insets show dispersion curves in the Brillouin zone near the $K$- and $K^{'}$-points.
The band inversion between the two different domains is characteristic of a valley-Hall topological waveguide.
(b) Propagation loss of acoustic waves as reported in the literature.
 Red, light blue, and deep blue indicate experiments at room temperature, at cryogenic temperatures above 4 K, and at the millikelvin (mK) level, respectively. Diamonds\cite{Xi2021, yu2020topo, Hatanaka2024} represent topological devices, while circles\cite{Hatanaka2014, Hatanaka2015, Cha2018, fu2019phononic, Patel2018, Romero2019, Mayor2021, xu2022, Feng2023, Chen2023_om, Zhang2024, Bicer2024} represent non-topological systems. Markers with or without filling indicate on-chip or bulk devices, respectively. 
\label{f:fig_1}}
\end{center}
\end{figure}

Phonons are the excitations of sound in a solid or liquid. 
Compared with photons or electrons, they feature slow propagation speed, strong confinement in a material, and immunity to electromagnetic radiation.
These properties are fuelling the prospect of chip-scale phononic circuits which route, store and process classical or quantum information in more compact, efficient, and robust ways\cite{Eggleton2019, Safavi2019, Wang2024}. 
Over the last decade, a growing community of researchers has thus investigated and refined the properties of the elementary building block of such circuitry: a low-loss waveguide for phonons.
Work towards this goal has covered a wide range of frequencies, ranging from the ultrasonic\cite{Hatanaka2014, Hatanaka2015, Cha2018, Romero2019, yu2020topo} (ca. 20~kHz--100~MHz) to the hypersonic\cite{Fang2016, Patel2018, fu2019phononic,Mayor2021, xu2022, Zivari2022,Chen2023_om, Feng2023, Zhang2024, Bicer2024} (ca. 100~MHz--10~GHz) domain.
However, propagation losses have remained at a relatively high level ($\gtrsim \mathrm{dB/cm}=10^5\,\mathrm{dB/km}$), see Fig.~\ref{f:fig_1}, limiting the range of applications.

Recently, these efforts have been enriched significantly by the uptake of ideas from topological physics.
Non-trivial topological properties of sound propagation in materials patterned with suitable artificial crystal structures have been investigated\cite{ma2019topological, Xue2022, Shah2024},
mirroring similar effects for light studied in topological photonics.\cite{Ozawa2019}
Here, analogies with quantum spin- and valley-Hall physics are of fundamental interest, in particular the existence of edge modes that guide excitations at the interface between two non-equivalent topological insulators\cite{WuHu15,khanikaev2017two,barik2018topological}.
A more practical motivation lies in the promise of reduced backscattering within such edge modes in the presence of certain perturbations such as sharp bends---a highly desirable feature for a waveguide especially in complex circuits.
Recent implementations have provided some empirical evidence for such protected transport\cite{yan2018chip, Cha2018Nature, jwma2021, Xi2021, Zhang2022, Hatanaka2024}, yet quantitative measurements of backscattering are surprisingly scarce.\cite{Ren2022topo}
Crucially, however, topological protection does not entail low losses. In fact,  existing topological phononic waveguides are lossy (Fig. \ref{f:fig_1}b), which has remained an open challenge\cite{Shah2024}. 

We address the above challenges for phononic waveguides by drawing on the recently achieved dramatic advances on diluting the dissipation of nanomechanical resonators using high stress\cite{Sementilli2022}.
In particular, we harness the method of soft clamping\cite{Tsaturyan2017}, here characterized by a smooth exponential mode confinement. 
It suppresses both radiation loss to the substrate and internal dissipation due to material bending, and has triggered an avalanche of work that culminated in mechanical resonators with quality-factors (Q) above 10 billion \cite{Ghadimi2018, Reetz2019, Bereyhi2022, Hoj2024, Beccari2022, Cupertino2024}.
The ultralow loss of these localized, non-propagating phononic modes has enabled many applications in quantum opto-\cite{Rossi2018, Mason2019, Chen2020, Fedoseev2021, Huang2024} and electro-mechanics\cite{Seis2022}, for force sensing and microscopy\cite{Halg2021} and as a mechanical memory\cite{Kristensen2024}. 

In this work, we show how the concepts of topological phonon transport and soft-clamping naturally coalesce:
transverse to their propagation direction, the topological edge modes penetrate evanescently into the bulk crystal, giving rise to the soft confinement that is at the heart of soft clamping\cite{Tsaturyan2017}.
In an suitably patterned high-stress membrane, we thus obtain ultrasonic waveguides with propagation loss down to 3~dB/km, many orders of magnitude lower than anything demonstrated at ultra- or hypersonic frequency to date (Fig.~\ref{f:fig_1}).
Remarkably, the loss is comparable to superconducting waveguides for microwaves\cite{Kurpiers2017}, and approaches that of optical fibers. 

The low loss also allows us to accurately quantify the very small amount of backscattering in a waveguide that includes sharp ($120\degree$) bends, by high-resolution ultrasound spectroscopy of a waveguide wrapped up into a triangular cavity.
We conclude that the bends dominate backscattering, with an average probability of only  $1.1\times10^{-4}$.
That is, about $99{.}99\%$ of the phonon energy flows around an individual bend. 
Such ultra-low loss and  backscattering open many new opportunities for classical and quantum phononic circuits and interconnects.

\section*{A phononic valley-Hall insulator in a pre-stressed membrane}

\begin{figure}
\begin{center}
\includegraphics[width=\linewidth]{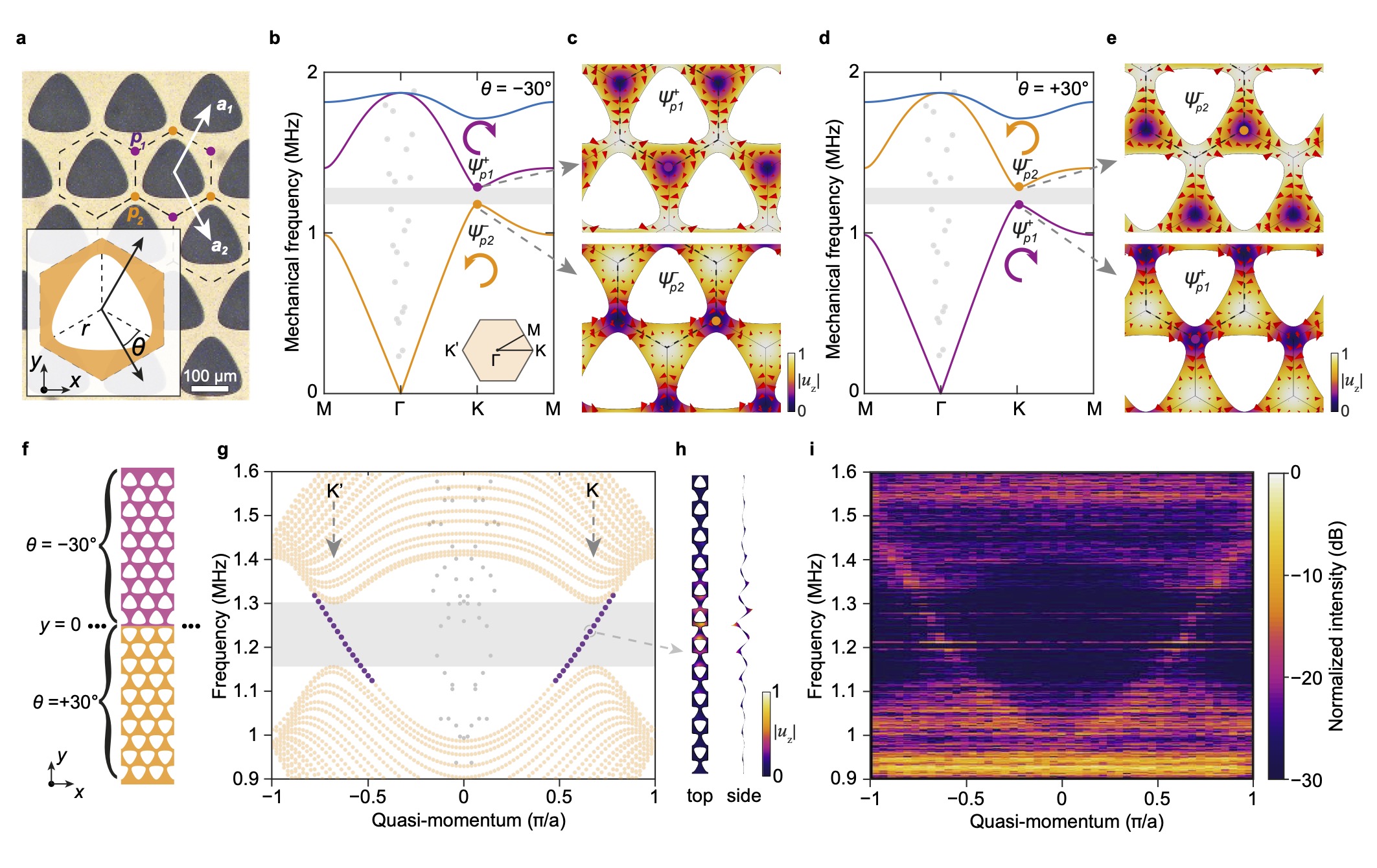}
\caption{{
\bf Valley-Hall topological insulator and topological edge states in a thin stressed membrane.}
(a) Microscope image of fabricated phononic valley-Hall crystals in a SiN membrane.
Superimposed are a hexagonal unit cell of the honeycomb lattice with the two inequivalent sites $p_1$ and $p_2$, as well as the basis vectors $\mathbf{a}_1$ and $\mathbf{a}_2$.
The inset shows a single unit cell with rounded triangular hole of circumradius $r$ and orientation $\theta$ in the centre.
(b, d) Simulated energy band diagrams for the crystals  with $\theta = -30 \degree$ (b) and $\theta = +30 \degree$ (d). 
At the $K$-point, the valley states $\psi_{p1,p2}^{\pm}$ have an clockwise ($+$) or anticlockwise ($-$) energy flow, with vortices centered at sites $p1$ and $p2$. 
(c, e) Simulated modal profiles of the states $\psi_{p1}^{+}$ and $\psi_{p2}^{-}$ in (b) and (d). %
The colour code represents $u_z$, the out-of-plane mechanical field and red arrows represent the energy flow. 
(f) Design of the domain interface that accommodates topological edge states. 
(g) Simulated band diagram of the topological structure in (f). The yellow, purple, and grey dots represent bulk modes, topological edge modes, and in-plane mechanical modes, respectively.
Additional boundary modes localised at the membrane's rims are not presented.
(h) Top and side views of simulated spatial distributions of the topological edge states at the $K$-point.
(i) Measured band diagram of the topological insulator along the topological domain interface. 
\label{f:fig_2}}
\end{center}
\end{figure}

In this work, we implement topological phononic waveguides in 20-nm thin silicon nitride (SiN) membranes under high tensile stress ($\sim1.2~\mathrm{GPa}$), as required for dissipation dilution.\cite{Schmid2023}
The phonon modes of interest are flexural vibrations out of the membrane plane (Fig.~\ref{f:fig_1}a).
We then adapt ideas that have been used successfully to demonstrate topological modes in linear media\cite{Lu2017,yan2018chip,jwma2021, Ren2022topo, Zhang2022} to a stressed membrane. 
In particular, we implement an analogue of the valley-Hall effect, by patterning the membrane with a periodic array of holes.
The holes are arranged on a honeycomb lattice whose basis vectors $\mathbf{a}_1$ and $\mathbf{a}_2$  are $a\coloneqq \SI{200}{\micro\meter}$ long (Fig.~\ref{f:fig_2}a).
Each hole has the shape of a rounded triangle with a circumradius $r<\SI{100}{\micro\meter}$.
The triangle's symmetry axis and a honeycomb lattice vector subtend an angle $\theta$ (Fig.~\ref{f:fig_2}a).

This geometry implies a rotational symmetry $C_3$.
For $\theta = 0 \degree$,  an additional mirror symmetry 
yields the point group $C_{3v}$.\cite{kovsata2021second}
In phonon momentum space, doubly-degenerate Dirac cones then form at the $K$ and $K'$-points on the edge of the first Brillouin zone, similar to graphene-like structures\cite{Lu2014_prb_Dirac_cones}. 
For $\theta\neq0\degree$, the mirror symmetry is broken.
This reduces the corresponding point group to $C_3$ and simultaneously lifts the double degeneracy to create a bandgap (Figs.~\ref{f:fig_2}b,d,S5).
In the vicinity of the $K$-point, the local band structure can be modelled with a perturbative Dirac-Hamiltonian
$     H =  v_{\mathrm{D}}(\delta k_x \tau_2 + \delta k_y \tau_1) + m \tau_3$, 
where $v_{\mathrm{D}}$ is an effective group velocity, $\delta k_x$ and $ \delta k_y$ are the quasimomenta relative to $K$, $m$ is an effective mass term, and the $\tau_i$ are the Pauli matrices. 
The corresponding eigenvalues $E_\pm=\pm [v_\mathrm{D}^2(\delta k_x^2 + \delta k_y^2)+m^2]^{1/2}$ form hyperbolid sheets centered at the $K$-point, separated by $2m$---which therefore quantifies the bandgap.
Time-reversal symmetry implies that a similar band structure forms at the $K'$-point  (see inset Fig.~\ref{f:fig_1}a).
In the following, we are predominantly concerned with the propagation of phonons whose wave vector is located in one of these so-called valleys, i.e., close to the $K$ and $K'$ points.

At the $K$-point, the associated eigenfunctions $\psi_{p1}^{+}$ and $\psi_{p2}^{-}$ feature vortices with opposite chirality centered at two inequivalent triangular-lattice sites $p_1$ and $p_2$ in real space (see Fig. \ref{f:fig_2}b-e).
Crucially to our work, inverting the triangle orientation  $\theta\rightarrow -\theta$ leaves the band structure intact, but inverts the symmetry-breaking perturbation, i.e., transforms $+m \rightarrow -m$.
This mass inversion enforces a  band inversion, switching the vorticity of the valley states (Fig.~\ref{f:fig_2}b-e).
The two bulk crystals with opposite $m$ are topologically distinct: they cannot be continuously deformed into one another without closing the gap. 

In order to obtain a localized topological edge mode, we construct  at a zigzag interface between two bulks with $\theta = +30\degree$ and $\theta = -30\degree$, respectively (Fig.~\ref{f:fig_2}f). 
The changing sign of the mass term $m$ across the interface entails a topological chiral edge mode within the gap at each valley, according to the bulk-boundary correspondence (Supplementary Information).
Modes propagating along the edge are hence located close to either the $K$- or $K'$-valleys, depending on their propagation direction.
Backscattering is expected to be suppressed due to large required transfer of quasimomentum, available only in structures that break the crystal symmetry.\cite{Shah2021, Shah2024}

In our implemented design, it is critical to properly account for the non-trivial relaxation of the pre-stress in the bulk as well as the interface region, upon patterning the membrane with holes. 
With a triangle size $r \approx \SI{93}{\micro\meter}$ that we choose in this work,
a numerically simulated band diagram shows a $\sim 105\,\mathrm{kHz}$-wide gap centred around  $1.23\,\mathrm{MHz}$ at the $K$ and $K'$ valleys (Fig.~\ref{f:fig_2}g).
Indeed across the gap, in each valley, we find one mode with a mid-gap group velocity of $|v_g| = 280~\mathrm{m/s}$.
Evaluating the associated spatial wavefunction, we confirm that the mode is localized to the interface (Fig.~\ref{f:fig_2}h).
Transverse to the propagation direction, it decays exponentially with a decay length $L_\mathrm{c}\sim \SI{280}{\micro\meter}$ (see Fig.~\ref{f:fig_2}h and Supplementary Information), thus providing soft mode confinement.

We fabricated devices according to this design (Supplementary Information). 
We map out the phonon band diagram by exciting the edge mode in one point and tracking its evolution along the edge (Supplementary Information).
The experimental results are shown in  Fig.~\ref{f:fig_2}i, and agree well with the simulated results.
This confirms the existence of phononic valley edge states in SiN membrane in presence of high stress.

\section*{Ultralow-loss phonon propagation}

The successful implementation of topological edge states in highly-stressed thin membranes sets an expectation of low propagation loss,  following the recently refined understanding that out-of-plane vibrations of thin membranes under high tensile stress benefit from dissipation dilution.\cite{Schmid2023,Sementilli2022}
Specifically, this implies that material parameters, such as the intrinsic quality factor $Q_\mathrm{int}=E_1/E_2$ (the ratio of the real and imaginary parts of the elastic modulus) do not exclusively determine the ultimate $Q$-factor of a mode (with mode index $n$).
Rather, it is enhanced to $Q_n=D_n\cdot Q_\mathrm{int}$ by the dissipation dilution factor $D_n\gg 1$, which benefits from higher stress, smaller thickness, and smooth mode confinement.
Physically, this is enabled by storing a large fraction $\sim D_n/(D_n+1)$ of the total mode energy in a loss-less potential $W^\mathrm{tens}_n$, namely the elongation of the material against the static tensile stress.
In contrast, $W^\mathrm{bend}_n$ is the energy stored in membrane bending, of which a fraction $\sim 2\pi /Q_\mathrm{int}$ is dissipated each cycle.
This leads to a dilution factor $D_n\approx W^\mathrm{tens}_n/W^\mathrm{bend}_n$ for strongly stressed membranes, and the imperative to engineer modes which minimize bending\cite{Schmid2023} (Supplementary Information).

Engineering mode geometry to follow this imperative is the basic idea of soft clamping.\cite{Tsaturyan2017}
Confining vibrational modes to defects in phononic crystals has proven extraordinarily effective in this regard, since the mode can penetrate evanescently into the crystal. 
It thereby ``softly'' transitions to low amplitudes, avoiding the sharp kinks resulting from the boundary conditions at a hard mechanical clamp.\cite{Schmid2023}
Given the exponential decay of the topological edge modes into the bulk, soft clamping would be expected in our high-stress devices. 

\begin{figure}
\begin{center}
\includegraphics[width=\linewidth]{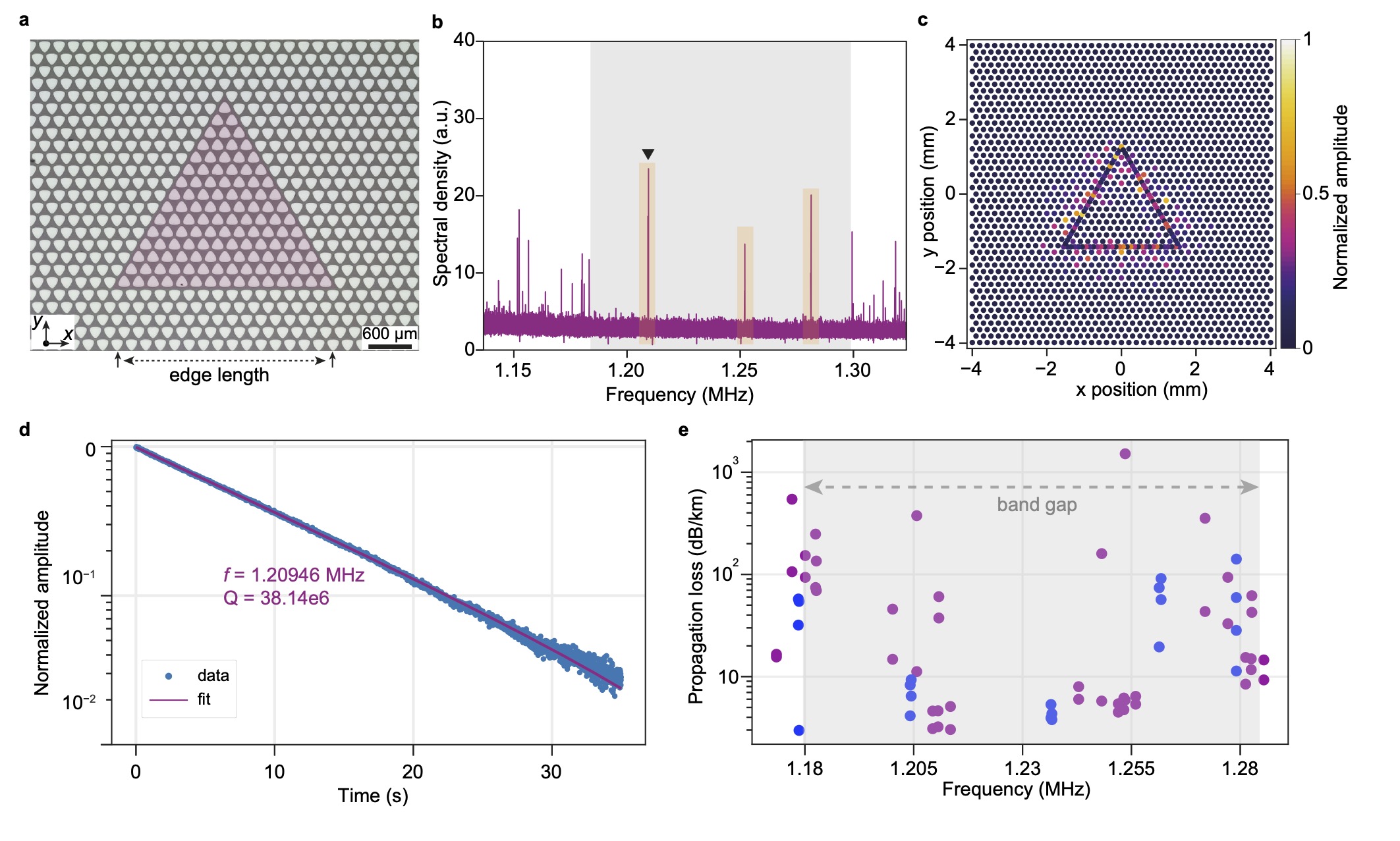}
\caption{{\bf Triangular-waveguide phonon cavities with ultra-low loss.} 
(a) Microscope image of a topological phonon waveguide wrapped up into a triangular cavity. Side length is 15 unit cells.
(b) Thermomechanical noise spectrum measured on a waveguide cavity as shown in (a). 
Gray shading indicates the bulk bandgap.
(c) Spatial map of thermomechanical noise amplitude at a cavity resonance frequency (marked with a triangle in (b)), confirming localisation to the domain interface. 
(d) Ring-down measurement of one of the cavity modes (marked with triangle in (b)). 
(e) Maximal propagation loss of edge modes inferred from measurements on several triangular cavities with edge length of 15 (purple) or 19 (blue) unit cells. 
The shaded region indicates an averaged bandgap, whose frequency may vary by 15 kHz from device to device.
\label{f:fig_3}}
\end{center}
\end{figure}

We thus anticipate that edge mode phonons propagate over distances much longer than any feasible chip-size, and
 assess the loss in a topological waveguide wrapped up into a closed triangular path (Fig.~\ref{f:fig_3}a).
Each side of the triangle is 15 (19) unit cells long in a first (second) batch of samples.
At the three corners, sharp $120^{\degree}$-bends connect the sides.
Importantly, the geometry of the bends largely conserve the crystal symmetry.
The distance of the triangular cavity to the membrane rim exceeds $\SI{2000}{\micro\meter}\gg L_\mathrm{c}$, which prompts us to neglect radiation loss in the following.

The additional periodic boundary condition imposed by the closed path along the triangular edge selects a set of discrete modes with longitudinal wave numbers $k^\parallel_n=2\pi n/L$, where $n$ is an integer and $L\approx 9{.}3~\mathrm{mm}$ is the round-trip length of the samples with 15 unit cells side length.
Through the waveguide dispersion (cf.\ Fig.~\ref{f:fig_2}g), the discrete set of allowed wavenumbers maps to a set of well-defined frequencies $\{\Omega_n\}$, at which an edge mode oscillates. 

Without external actuation, we measure a thermal spectrum (Fig.~\ref{f:fig_3}b) of the out-of-plane displacement at a location along the topological edge, with a sensitive laser interferometer at room temperature (Supplementary Information).
It reveals a series of sharp peaks close to the frequencies $\{\Omega_n\}$ inside an otherwise quiet region, corresponding to the band gap. 
(Peaks around any one $\Omega_n$ actually appear in pairs, the origin of which will be discussed below.)
More dense and irregular peaks are seen outside the bandgap. These correspond to bulk modes.
We also measured the spatial localization of these topological modes by scanning the laser spot over the membrane in a coarse grid matched to the phononic crystal, with one example shown in Fig.~\ref{f:fig_3}c.
At the edge mode frequencies, we find large motional amplitudes only close to the interface between the two bulk crystals, confirming the localisation of the topological edge states.

The propagation loss $\alpha$  
for waves guided along the edge sets a lower limit for the rate $\Gamma_n$, at which the energy of the phonons circulating in the triangular cavity is dissipated.
On this basis, we can relate measurements of the triangle-cavity's mode quality factors $Q_n=\Omega_n/\Gamma_n$ to the propagation loss in the waveguide via 
$\alpha_n\lesssim  \Omega_n/Q_n v_\mathrm{g}$.\cite{ Patel2018, fu2019phononic}
Here, we use the simulated mid-gap value $v_g = 280~\mathrm{m/s}$ for the group velocity, which we found to slightly underestimate the measured group velocities.
We use cavity ringdowns as a robust technique to determine the modes' $Q_n$ with time-domain measurements, with an example shown in Fig.~\ref{f:fig_3}d.
We observe a high quality factor of $Q_n=38{.}1\times 10^6$ for a cavity mode at $\Omega_n\approx2\pi\cdot 1{.}21~\mathrm{MHz}$, corresponding to a propagation loss of less than $3.1~\mathrm{dB/km}$.
This is a remarkably low loss, even more so for an on-chip waveguide.
Measuring the loss in many devices (Fig.~\ref{f:fig_3}e) we find some statistical scatter, similar to that of soft-clamped zero-dimensional resonators\cite{Tsaturyan2017}.
The modes near the edge or outside of the bandgap present systematically higher loss, likely because of hybridisation with modes at the membrane rim. 

Finite element modelling furthermore allow us to quantitatively evaluate the dilution factors $D_n$ for the topological edge modes, yielding values in the range of {10{,}000-25{,}000} for the different in-gap modes.
Together with the intrinsic quality factor\cite{Villanueva2014} {$Q_\mathrm{int}\approx 1140 $} for the 20-nm thick SiN membranes employed in this work, we obtain quality factors well compatible with our measurements (Fig.~\ref{f:fig_3} and Supplementary Information).
Simulations also confirm that hard-clamped waveguides in stressed membranes\cite{Romero2019} are limited to much higher propagation loss (Supplementary Information).

It is instructive to recast the triangular phonon cavity's quality factor as a finesse using $\mathcal{F}_n= Q_n \cdot \Omega_\mathrm{FSR}/\Omega_n$, with the free spectral range $\Omega_\mathrm{FSR}/2\pi= v_\mathrm{g}/L.$
For the mode with the highest $Q$-factor, we obtain a cavity finesse of $\mathcal{F}_n\approx 1.3\times 10^6$, on par with the highest finesse that can be realized for an optical cavity using dielectric mirrors.\cite{Rempe1992}
With the round-trip loss given by $\pi/\mathcal{F}$, we can further infer that when passing a sharp bend---of which there are three per round-trip---less than $\sim0{.}8\,\mathrm{ppm}$ of phonons are lost on average.
This is a stark illustration of the fact that phonons cannot be scattered to free-space modes, a common limitation in optical cavities\cite{Rosiek2023}, and that scattering to other modes in the device is strongly suppressed, not least by the absence of other modes in the gap.

\section*{Quantifying backscattering}

\begin{figure}
\begin{center}
\includegraphics[width=\linewidth]{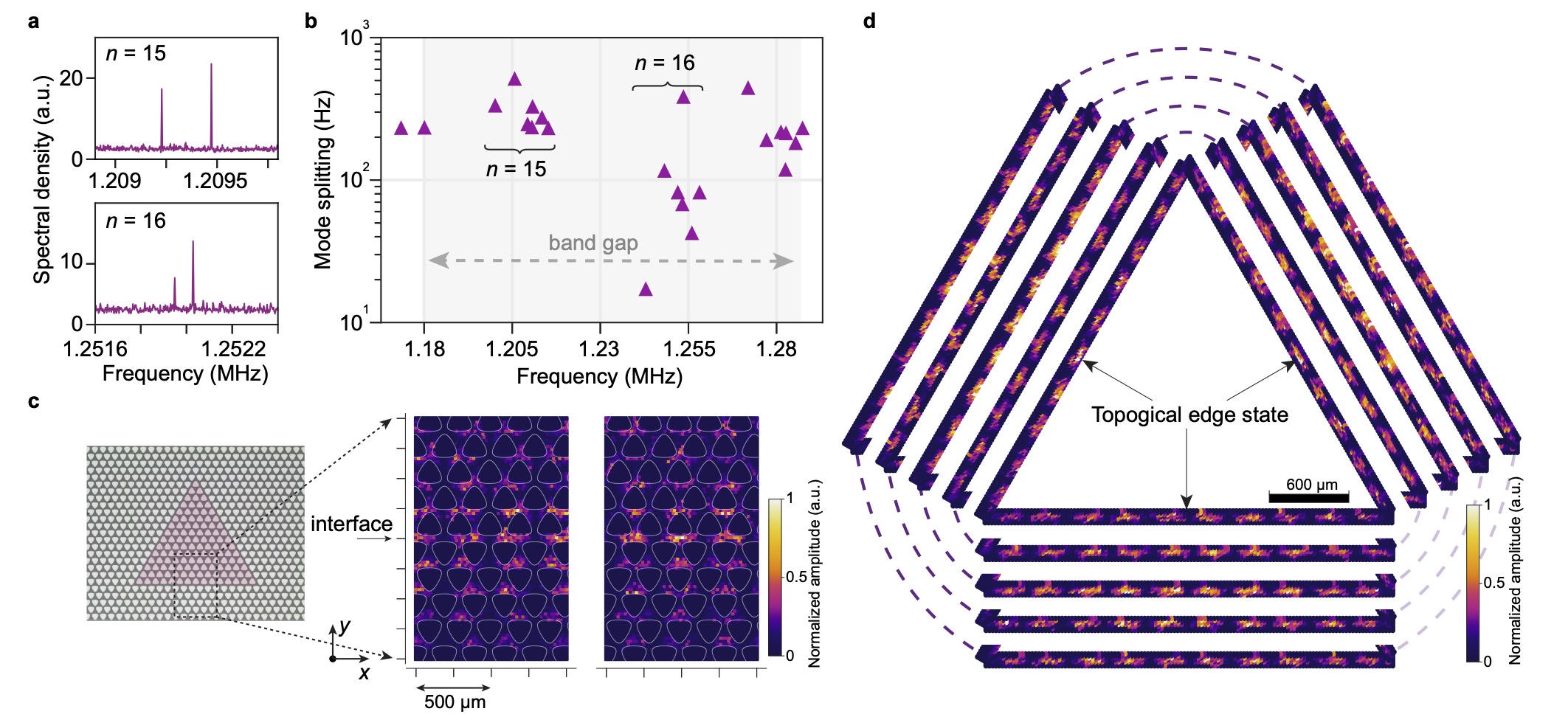}
\caption{{\bf Backscattering-induced mode splitting.}
(a) Zooms into the spectral density of Fig.~\ref{f:fig_3}b reveal small splittings for the cavity modes.
(b) Overview of measured mode splittings from different triangular cavities with edge length of 15 unit cells.
Devices are discarded if all edge modes have $Q_n<10^7$. 
(c) Spatial map of thermomechanical noise amplitude for the mode pair shown in the upper panel of (a). 
The position of nodes and anti-nodes indicates that the two modes are spatially orthogonal to each other. 
(d) Spatial maps of thermomechanical noise amplitude of one of the  $n = 15$ edge modes along the cavity's waveguide path.
The shown maps from five different devices align, suggesting consistent location of scatterers, namely the corners.
\label{f:fig_4}}
\end{center}
\end{figure}

Closer inspection of the measured and simulated spectra reveals that the in-gap modes occur in pairs (Fig. ~\ref{f:fig_4}a), whose frequencies $\Omega_n^\pm$ are separated by a small splitting $\Delta\Omega_n\equiv(\Omega_n^{+}-\Omega_n^{-})\approx{ \mathcal{O}(2\pi\cdot 100~\mathrm{Hz})}$.
Due to time-reversal symmetry, clockwise and counter-clockwise propagating edge modes are expected to be degenerate (see also Fig.~\ref{f:fig_2}g).
However, residual coupling between counter-propagating modes due to backscattering will hybridise the travelling waves into standing waves, with their degeneracy lifted by the coupling.\cite{Shah2021, Ren2022topo}

The theoretical description of valley-Hall topological systems predicates that the counter-propagating waveguide modes are localized in the inequivalent $K$ and $K'$ valleys. 
The required large quasi-momentum transfer is therefore expected to suppress elastic backscattering---even at sharp bends as long as they preserve crystal symmetry sufficiently well.\cite{Shah2021, Shah2024}
Indeed, the theoretically expected robust transport along the edge channels of valley-Hall bosonic systems has provided a strong motivation for exploring such systems in the first place\cite{Hafezi2011}.
Meanwhile, numerous experiments both in the optical and acoustic domain have qualitatively established guidance in topological edge modes around sharp bends.
However, experiments that accurately quantify residual backscattering are very rare---and when conducted, have sometimes yielded seemingly unexpected results\cite{Rosiek2023}.
To our knowledge, the most precise characterization for topological phonon waveguides was obtained recently by a qualitative comparison of simulated and measured cavity spectra, from which a reflection probability $p_\mathrm{bs}<5\%$ upon passage of a sharp corner was inferred.\cite{Ren2022topo} 

In our devices, the low loss and associated narrow linewidths $ \Omega_n^\pm/Q_n^\pm\ll2\pi\cdot1~\mathrm{Hz}$ afford precise spectroscopic resolution of the splittings across a large number of cavity resonances and devices, as shown in Fig.~\ref{f:fig_4}b.
The splitting is proportional to the backscattering amplitude, and correspondingly its probability to occur upon passage of a single (effective) scatterer in a cavity round-trip can be given as $p_\mathrm{bs}=(\pi \cdot \Delta\Omega_{n}/\Omega_\mathrm{FSR})^2$ (Supplementary Information).
For the smallest measured splitting  $\Delta\Omega_n=2\pi\cdot17.2~\mathrm{Hz}$, this evaluates to a backscattering probability of $p_\mathrm{bs}^\mathrm{min}\approx2\cdot 10^{-6}$ in the whole device, however the values vary strongly among modes and devices.

To refine the understanding of the origin of backscattering, we complement the spectroscopic analysis with spatial scans of the mode patterns.
To this end, we map out the amplitude of thermal excitation over a finer grid of spatial sampling points 
along the interface forming the triangular cavity (Fig.~\ref{f:fig_4}).
We obtain regular patterns well consistent with standing waves in the cavity.
Their respective number of antinodes ($2n$), first of all, allows straightforward assignment of mode numbers to frequencies as $\{\Omega_{n=15},\Omega_{n=16},\Omega_{n=17}\}\approx2\pi\{1.210, 1.255,1.280\}~\mathrm{MHz}$.
Zooming further into a straight waveguide region (Fig.~\ref{f:fig_4}c) for patterns of the $\Omega_{n=15}^\pm$ mode pair, reveals that the two standing wave patterns are orthogonal to each other, with antinodes of one mode lining up with the other mode's nodes. This would be expected for symmetric and anti-symmetric combinations of counterpropagating Bloch waves.
Furthermore, we find that the locations of nodes and antinodes are consistent among a series of five samples 
(Fig.~\ref{f:fig_4}d). 
This supports the notion of scatterers that are reproduced in our fabrication process, and hence likely part of the mask geometry, rather than random disorder introduced by fabrication imperfections or contamination, which could lead to Anderson localization\cite{Rosiek2023}.
We also observe that the $\Omega_{n=15}^\pm$  modes have either nodes or antinodes  pinned to the corners of the triangular phonon cavity.
Together, we take this evidence to suggest that backscattering occurs mainly at the three corners of the triangle.

Under this assumption, it can be shown that, due to interference, the backscattered waves from the corners add up for modes with $(n \mod 3)=0$, whereas otherwise backscattering cancels out for identical scattering amplitude from each corner (Supplementary Information).
In real devices, such cancellation is always imperfect, which could well explain the larger variation, but also particularly low splitting, observed for the $n=16$ mode pair.
Following this line of reasoning, we can extract the mean splitting $\Delta\Omega_{n=15}/3$ caused by the passage of one corner. 
Averaging over all measured samples, we obtain $\langle\Delta\Omega_{n=15}\rangle/3\approx2\pi\cdot 102\,\mathrm{Hz}$, corresponding to a backscattering probability $p_\mathrm{bs}^{(\angle)}=1{.}1\cdot10^{-4}$ upon passage of an individual corner.
We suspect the backscattering comes from the slightly broken symmetry due to the different stress distribution at the corner.

As a benchmark, we have also simulated a triangular cavity made from a hard-clamped membrane waveguide. 
We obtain dramatically larger splittings for modes with $(n \mod 3)=0$, comparable with the cavity free spectral range, as would be expected for a backscattering probability of order unity at a corner (Supplementary Information).
This underlines the strong backscattering suppression achieved in our topological phonon waveguides.

\section*{Parametric amplification}

To date, most of the topological waveguiding systems remain working in the linear regime.
However, the nanomechanical devices introduced here also give access to richer physics that occurs in non-linear and time-modulated potentials.
This includes the Duffing nonlinearity and squeezing interactions, which are well understood in this platform\cite{Catalini2021,Schmid2023, Xiang2024}.
The low loss of  our topological phononic waveguides also offer opportunities for exploring the nonlinear phenomena with topological physics. 
As an experimental example, we  subject one of the triangle cavities to a parametric drive, which modulates membrane stress and thereby the resonance frequencies  
$\Omega_{n}^\pm \rightarrow \Omega_{n}^\pm (1+\epsilon\cos(\Omega_\mathrm{p} t))$.
Tuning $\Omega_\mathrm{p}$ to twice the frequency of one of the topological edge modes, we observe phase-dependent amplification of that mode with a small mechanical pump (Supplementary Information).

\section*{Conclusions and outlook}
In conclusion, we have realized a chip-based platform that unites topological phonon waveguiding and dissipation dilution. 
The resulting devices achieve unprecedentedly low propagation loss, and very low backscattering even at sharp bends.
Our analysis suggests that whereas the observed quality factors are well above the hard-clamp limit, there is room for further optimization of the dilution factor (Supplementary Information). 
We therefore expect that quality factors and propagation losses can be improved significantly, similar to the exponential gains seen recently in the Q-factors of soft-clamped localized resonators\cite{Engelsen2024}.
Backscattering, in turn, may be reduced further by engineering the interface between the bulk crystals\cite{Shah2021}.
Such a platform suggests itself for routing, processing and storage of signals, in which linear elements such as waveguides, resonators, and directional couplers are combined to a complex circuit with ultralow power dissipation. 

The intrinsic nonlinearity and the possibility to locally change the mechanical potential electrostatically\cite{Halg2022} add a host of opportunities for non-Hermitian topological physics, which we have only begun to explore. 
Single electrodes could implement phase shifters or tunable parametric mode-couplers\cite{Halg2022}.
Electrode arrays with phase-synchronized drives could provide an architecture for low-noise travelling wave amplification\cite{Peano2016}, phononic solitions, and so on.

Applications for quantum signals become of interest when the devices are cooled even  to only moderate cryogenic temperatures, at which the  material’s $Q_\mathrm{int}$ typically increases five-fold\cite{Rossi2018}.
We estimate the length that a quantum signal can travel before an incoherent phonon leaks into the waveguide as $\Lambda\approx 1/ n_\mathrm{th} \alpha=v_\mathrm{g}\cdot \hbar Q/k_\mathrm{B} T$, yielding $\sim 10 \,\mathrm{cm}$ ($8\,\mathrm{m}$) at a temperature $T$ of $ 4 \,\mathrm{K}$ ($ 50 \,\mathrm{mK}$) of the thermal environment ($k_\mathrm{B}$ and $\hbar$ are Boltzmann and Planck's reduced constant, respectively)---ample range to implement even complex phononic circuits on a chip. 
Quantum-coherent interfaces between soft-clamped phononic resonators and light\cite{Rossi2018} as well as superconducting circuits\cite{Seis2022} have already been demonstrated, and are under investigation with single spin systems\cite{Fung2023,Hahne2023}. 
Combination with the here-demonstrated topologically protected phononic circuits thereby only advances the prospects for information processing in hybrid quantum systems.\cite{Kurizki2015}

\paragraph*{Acknowledgements}
The authors would like to acknowledge help with micro-fabrication by Dr.~Thibault Capelle.
This work was supported by the European Research Council project PHOQS (Grant No. 101002179), the Novo Nordisk Foundation (Grant Nos. NNF20OC0061866 and NNF22OC0077964), the Danish National Research Foundation (Center of Excellence “Hy-Q”), the Independent Research Fund Denmark (Grant No. 1026-00345B), the Swiss National Science Foundation (CRSII5 177198/1, CRSII5 206008/1, PP00P2 163818), the Deutsche Forschungsgemeinschaft (DFG) via project number 449653034 and through SFB1432, European Union's Horizon 2020 research and innovation programme under the Marie Skłodowska-Curie grant agreement No. 101107341, and a research grant (VIL59143) from the Villum Foundation .

\paragraph*{Competing interests}
The authors declare no competing interests.

\paragraph*{Author contributions}

X.X. designed the devices and built the setup with contributions from I.C..
X.X. and I.C. conducted the measurements and analyzed the data. 
J.K. and O.Z. contributed to the design at early stage and developed the topological theory.
X. X., J. K., M. B. K. and E. L. conducted numerical simulations.
X.X, M.B.K. and A.S. analysed dissipation dilution and soft clamping. 
E.L. fabricated the devices. 
A. S. S. contributed to the discussion and understanding of  the experimental data.
X.X., J.K., and A.S. wrote the paper with the input from all the authors. 
X.X. and A.S. supervised the project.

\clearpage

\bibliography{topological.bib}

\newpage

%%%%%%%%%%%%%%%%%%%%%%%%%%%%%%%%%%%%%%%%%%%%%%%%%%%%%%%%%%%%%%%%%%%%%%%%%%%%%%%%%%%%%
%%%%%%%%%%%%%%%%%%%%%%%%%%%%%%%%%%%%%%%%%%%%%%%%%%%%%%%%%%%%%%%%%%%%%%%%%%%%%%%%%%%%%
%%%%%%%%%%%%%%%%%%%%%%%%%%%%%%%%%%%%%%%%%%%%%%%%%%%%%%%%%%%%%%%%%%%%%%%%%%%%%%%%%%%%%
%%%%%%%%%%%%%%%%%%%%%%%%%%%%%%%%%%%%%%%%%%%%%%%%%%%%%%%%%%%%%%%%%%%%%%%%%%%%%%%%%%%%%
%%%%%%%%%%%%%%%%%%%%%%%%%%%%%%%%%%%%%%%%%%%%%%%%%%%%%%%%%%%%%%%%%%%%%%%%%%%%%%%%%%%%%
%%%%%%%%%%%%%%%%%%%%%%%%%%%%%%%%%%%%%%%%%%%%%%%%%%%%%%%%%%%%%%%%%%%%%%%%%%%%%%%%%%%%%
%%%%%%%%%%%%%%%%%%%%%%%%%%%%%%%%%%%%%%%%%%%%%%%%%%%%%%%%%%%%%%%%%%%%%%%%%%%%%%%%%%%%%
%%%%%%%%%%%%%%%%%%%%%%%%%%%%%%%%%%%%%%%%%%%%%%%%%%%%%%%%%%%%%%%%%%%%%%%%%%%%%%%%%%%%%

\newpage

%%%%%%%%%%%%%%%%%%%%%%%%%%%%%%%%%%%%%%%%
%%%                                                                                                          %%%%
%%%                     SUPPLEMENTARY INFORMATION                            %%%%
%%%                                                                                                          %%%%
%%%%%%%%%%%%%%%%%%%%%%%%%%%%%%%%%%%%%%%%
\renewcommand{\figurename}{{\bf Supplementary Fig.}}
\renewcommand{\tablename}{{\bf Supplementary Table}}
\setcounter{figure}{0}\renewcommand{\thefigure}{{\bf S\arabic{figure}}}
\setcounter{table}{0}\renewcommand{\thetable}{{\bf S\arabic{table}}}
\setcounter{equation}{0}\renewcommand{\theequation}{S\arabic{equation}}
\newgeometry{top=40mm, bottom=20mm, right=20mm, left=20mm}

\vspace{3mm}

\hspace{-6mm}{\LARGE \bf Supplementary Information}

\vspace{5mm}

\tableofcontents

\newpage

\section{Device fabrication}
The membranes were fabricated from stoichiometric silicon nitride thin film deposited on standard silicon wafers by low pressure chemical vapor deposition (LPCVD) with a resulting tensile stress of approximately 1.2 GPa. First, we spin-coated $\SI{1.5}{\micro\meter}$ thickness of resist AZ Mir 701 on both sides of the wafer. Second, the resist on the polished side of the wafer was patterned with phononic crystals, and the resist on the other side was patterned with the backside etch windows, by using a Heidelberg MLA150 maskless lithography system. After the development of photolithography, both sides of the wafer were then dry-etched with inductively coupled plasma (ICP) with $\mathrm{CF}_4/\mathrm{H}_2$ chemistry, transferring the pattern on the resist to the silicon nitride layer. Next, the membranes were released by etching the silicon with KOH wet chemistry, and a backside protective holder manufactured by Advanced Micromachining Tools (AMMT) was used to protect the phononic crystal side. Lastly, the membranes were cleaned using an $\mathrm{H}_2\mathrm{SO}_4/\mathrm{H}_2\mathrm{O}_2$ mixture for 10 minutes, followed by a rinse in deionized water, then allowed to dry in air. 

\section{Experimental setup}

\begin{figure}[ht]
\begin{center}
    \includegraphics[width=0.7\linewidth]{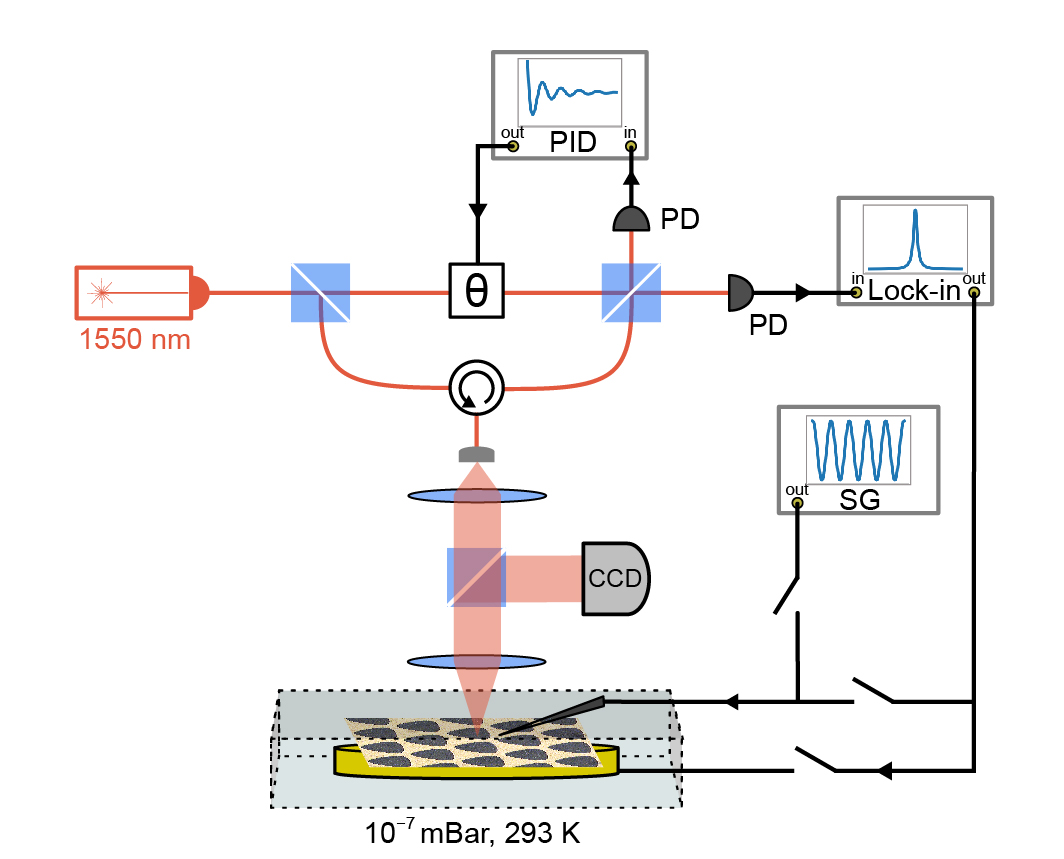}
    \caption{{\bf Experimental setup.} Shown is the fiber laser (1550 nm), the two photodetectors (PD), the proportional–integral–derivative (PID) controller  (FPGA: Red Pitaya), a phase shifter ($\mathrm{\theta}$, piezo fiber stretcher), the lock-in amplifier (lock-in), and a signal generator (SG). 
    The membrane is simultaneously imaged with a CCD camera.
    }
    \label{f:subFig_setup}
\end{center}
\end{figure}

Figure \ref{f:subFig_setup} shows the experimental setup for characterizing the devices. The devices were placed inside a vacuum chamber with a pressure below $10^{-6}$ mbar at room temperature. A home-built fiber-based Mach-Zender optical interferometer with the probe laser mounted on a programmable motorized stage was used to detect the mechanical motion of the devices across different spatial points.\cite{Barg2017} 

In the interferometer, the output of a 1550-nm fiber laser is split into a reference beam and a detection beam. The detection beam is focused on the membrane, and its reflection carrying mechanical motion signal is collected and then mixed with the reference beam via a beam splitter. The waist diameter of the focused beam on membrane is around $\SI{50}{\micro\meter}$. The outputs of the beam splitter are recorded by two photodetectors. One of them is connected to PID controller, whose output is connected to phase shifter to stabilize the relevant phase between the reference beam and detection beam at maximized detection efficiency point. Another one is connected to a Lock-in Amplifier and was used for mechanical signal analysis in both time domain and frequency domain.

The spectral dispersion diagram in {Fig.~2i} was obtained by measuring the mechanical frequency response to a micro-pin drive, while moving the detection laser spot along the topological interface. The micro-pin is gently brought into contact with the membrane near the topological interface, and it is also clamped to a piezoelectric actuator, which is driven by the lock-in amplifier. The presence of the pin strongly increases the mechanical loss, and it is therefore lifted off the surface for the other measurements. The spatial sampling rate of the laser spot is 8 points per unit-cell length. The band diagram is obtained by making spatial Fourier transform of the frequency response along the topological domain interface.

The spatial modal profiles was obtained by measuring the thermal spectra of the device across spatial points. In the measurement of the whole device ({Fig.3c}), we took one sampling point per lattice site in the bulk crystal region while we took three sampling points per unit-cell length along the topological interface line. We took a regular rectangular spatial grid along the selected region with sampling grid of $\SI{20}{\micro\meter}\times \SI{20}{\micro\meter}$ in {Figs.4c and \ref{f:corner_scan}}, and rectangular grid along the edge with mesh of $\SI{25}{\micro\meter} \times \SI{15}{\micro\meter} $ in {Figs.4d and \ref{f:tri_edge_scan_antinode}}.

For ringdown measurement of the mechanical modes, we use a piezo-electric transducer to shake the sample as a whole, at a frequency which we carefully tune to match one of the mechanical resonance $\Omega_n$.
Once we observe large-amplitude oscillations of one of the modes through the laser interferometer, the excitation is stopped.
The amplitude then decays as $\propto\exp(-\Gamma_n t/2)$ with time $t$, which allows us to extract $\Gamma_n$ from a fit and therefore $Q_n$.

\section{Theory of the topological phononic insulator} 

\subsection{Model of the bulk Hamiltonian, Berry curvature and Valley Chern numbers}

We briefly recount the arguments underlying the degeneracies of the band structure, their breaking and the appearance of edge modes. Similar treatment in more detail may be found in the literature~\cite{Saba2020, kovsata2021second, JanThesis}.

To derive the Hamiltonian of the bands near the $K$ and $K'$ points, we must find the symmetry-allowed perturbation terms. The band structure can be assumed to arise from the eigenvalue equation
\begin{equation} \label{seq:eigvaleq}
H_{\vb{k}} \boldsymbol{\Psi}_{\vb{k}}(\vb{r}) = \omega_{\vb{k}} \boldsymbol{\Psi}_{\vb{k}}(\vb{r}) \,,
\end{equation}
where $H_{\vb{k}}$ encodes the dependence on the structure geometry, including the pre-stress distribution. Since we assume the pre-stress to have the same symmetries as the lattice, we do not need to treat it explicitly.

Let us now choose a basis to express $H_{\vb{k}}$ at $K$ and $K'$. It can be shown~\cite{Bradley_2009} that the highest permitted degeneracy is fourfold, exemplified by the basis functions
\begin{equation} \label{seq:basis}
    \{ p_x(\vb{r}) e^{i \boldsymbol{K} \vdot \boldsymbol{r}}, p_y(\vb{r}) e^{i \boldsymbol{K} \vdot \boldsymbol{r}}, p_x(\vb{r}) e^{i \boldsymbol{K'} \vdot \boldsymbol{r}}, p_y (\vb{r})e^{i \boldsymbol{K'} \vdot \boldsymbol{r}} \} \,,
\end{equation}
where $p_x(\vb{r})$ and $p_y(\vb{r})$ are functions that transform as $x$ and $y$, respectively, and obey the lattice periodicity, i.e.,
\begin{equation}
    p_x(\vb{r} + \vb{a}) = p_x(\vb{r}) \,, \quad   p_y(\vb{r} + \vb{a}) = p_y(\vb{r})
\end{equation}
for any lattice vector $\vb{a}$.

We may now express the effects of the various symmetry operations in this basis. For the rotation $C_3$, reflection $\sigma_x$, inversion $I$, lattice translation $\vb{a}$ and time reversal $\mathcal{T}$, these read
\begin{equation}
\begin{aligned}
    \rho(C_3) &= \mathbb{1}_2 \otimes R_3\\
    \rho(\sigma_x) &= \mathbb{1}_2 \otimes \tau_3\\
    \rho(I) &= \tau_1 \otimes \mathbb{1}_2 \\
    \rho(\vb{a}) &= \text{diag} \left( e^{4\pi i  / 3}, e^{2\pi i  / 3} \right) \otimes \mathbb{1}_2 \\
    \rho(\mathcal{T}) &= \left(\tau_1 \otimes \mathbb{1}_2 \right)\mathcal{K}
\end{aligned}
\end{equation}
where $R_3$ is the $2\pi/3$ rotation matrix and $\mathcal{K}$ is complex conjugation.

The requirement placed on the Hamiltonian is that it commutes with the applicable symmetry operations, that is, for every present symmetry $g$,
\begin{equation} \label{seq:pt0}
H = \rho^{-1}(g) H \rho(g).
\end{equation}

For $\theta=0$, the applicable symmetries are $C_3$, $\sigma_x$, $\vb{a}$ and $\mathcal{T}$. It can be confirmed that the only Hamiltonian satisfying Eq.~\eqref{seq:pt0} is
\begin{equation}
    \qquad H = \omega_0 \mathbb{1}_4 \,,
\end{equation}
with a constant $\omega_0$. This accounts for the fourfold degeneracies shown in Fig.~\ref{f:bulk_band_simulation}a, with one band touching occuring at each $K$ and $K'$.

For $\theta \neq 0$, the reflection $\sigma_x$ no longer applies, allowing for
\begin{equation}
    H = \omega_0 \mathbb{1}_4 + m  \tau_3 \otimes \tau_2 \,,
\end{equation}
and thus opening a gap proportional to $m$ at $K$ and $K'$. Crucially, we see that the gap-opening term anticommutes with $\rho(I)$,
\begin{equation}
\rho(I)^{-1} \left( \omega_0 \mathbb{1}_4 + m \tau_3 \otimes \tau_2 \right) \rho(I) =  \omega_0\mathbb{1}_4 - m \tau_3 \otimes \tau_2  \,,
\end{equation}
that is, applying spatial inversion is equivalent to taking $m \, \rightarrow \, -m$. The two bulks thus created are said to be topologically distinct. This can be confirmed by integrating the Berry curvature around $K$ and $K'$ to obtain different valley Chern numbers for $m<0$ and $m>0$; comprehensive treatments of this formalism can be found in the literature~\cite{bernevig2013topological, shen2012topological}.

Evaluating the effect of moving away from $K$ and $K'$ is slightly more involved; we merely quote the result here and arrive at the complete four-band Hamiltonian
\begin{equation} \label{seq:Hfin}
    H = \omega_0 \mathbb{1}_4 + v_D \left( k_x \tau_3 \otimes \tau_1 + k_y \tau_3 \otimes \tau_2 \right) + m \tau_3 \otimes \tau_2 \,,
\end{equation}
where we interpret $v_D$ as the group velocity.
Note that $H$ may be viewed as two decoupled $2\cross2$ blocks, related by time-reversal symmetry. To simplify the following analysis, we will only consider one of these at a time, that is,
\begin{equation} \label{seq:Hred}
    H = \omega_0 \mathbb{1}_2 \pm v_D \left( k_x \tau_1 + k_y \tau_3 + m \tau_2 \right) \,.
\end{equation}
Finally, it is convenient to use a basis composed of the eigenstates of the mass term $\tau_2$, where $H$ reads
\begin{equation} \label{seq:Hredfin}
    H = \omega_0 \mathbb{1}_2 \pm v_D \left( k_x \tau_2 + k_y \tau_1 + m \tau_3  \right) \,.
\end{equation}
This is the Hamiltonian quoted in the main manuscript (where we use $\delta k_x$ and $\delta k_y$  for clarity in the context of the main manuscript).

Note that the basis we chose in Eq.~\eqref{seq:basis} corresponds to an irreducible representation of the underlying space group. An irreducible representation describes one possible symmetry of the solutions that may occur in the band structure -- we have thus shown that \textit{some} of the bands will obey Eq.~\eqref{seq:Hredfin}. The particular choice of basis is unimportant for this conclusion.

\subsection{Derivation of the edge states}

We now proceed to show the existence of a localised state at a domain wall by considering a space-dependent mass term $m \equiv m(x) = m \,\text{sgn}(x)$. As this breaks translation symmetry in the $x$ direction, $k_x$ is no longer a meaningful quantity - we must work in real space, where $k_x \rightarrow i \partial_x$. Taking for simplicity $k_y = 0$, Eq.~\eqref{seq:Hfin} appears as
\begin{equation}
    H = \omega_0 \mathbb{1}_2 \pm i v_D \tau_2 \partial_x+m \,\text{sgn}(x) \tau_3  \,.
\end{equation}
We may now solve the eigenvalue equation \eqref{seq:eigvaleq} to obtain $\boldsymbol{\Psi}(x)$. Requiring the solutions to vanish at $\abs{x} \rightarrow \infty$, we arrive at
\begin{equation}
    \boldsymbol{\Psi}(x) = e^{-m \abs{x} / v_D} \vb{\Psi}(0) \,,
\end{equation}
where $\vb{\Psi}(0) = (1,1)$. This solution, exponentially localised at the domain wall, is known as the Jackiw-Rebbi soliton~\cite{Jackiw_1976}. We note the localization is given by $m / v_D$ and is therefore dependent on both the gap size and the original band structure.

\subsection{Numerical simulation of bulk states}

\begin{figure}[hp] 
\begin{center}
    \includegraphics[width=\linewidth]{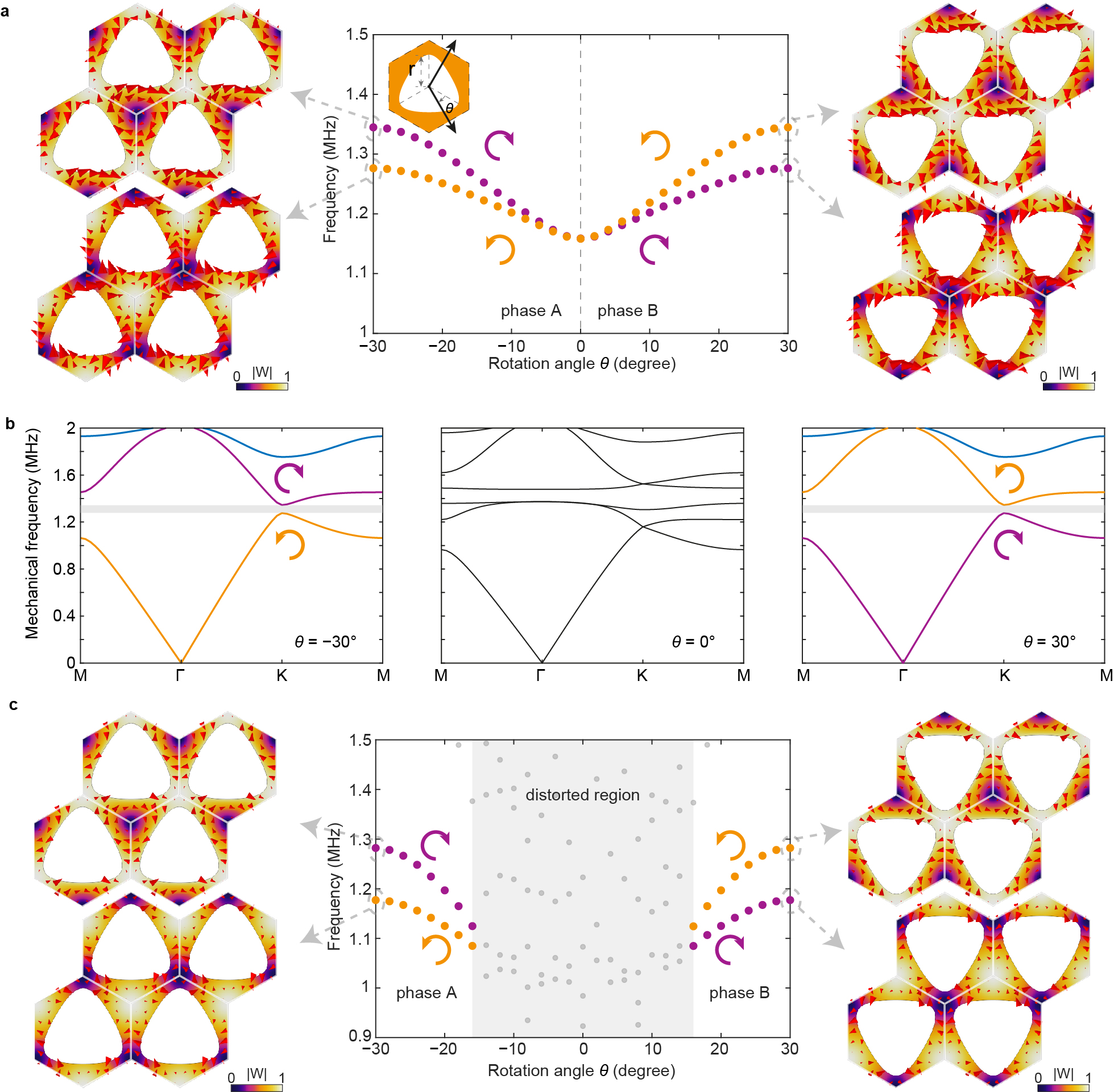}
    \caption{{\bf Band diagram simulation of bulk crystals.} (a) Eigenfrequencies above and below the topological bandgap at the $K$ point for different orientation $\theta$, with $r\approx\SI{85.3}{\micro\meter}$. Purple (yellow) dots denote the eigenfrequencies for flexural modes with clockwise (counter-clockwise) valley vortex. The flexural mode shapes show the displacement $\left|u_z\right|$ for $\theta = -30 \degree$ (left) and $\theta = 30 \degree$ (right), and the red arrows indicate the energy flow $\mathbf{P}(\mathbf{r}) \propto -iu_z^*(\mathbf{r}) \cdot \nabla u_z(\mathbf{r})$. (b) Simulated energy band diagrams for the crystal ($r\approx\SI{85.3}{\micro\meter}$) with $\theta = -30 \degree$ (left), $0 \degree$ (center), and $30 \degree$ (right). The bandgap is open at $K (K')$ point for a none-zero $\theta$ and closed for $\theta = 0 \degree$. (c)  Eigenfrequencies above and below the topological bandgap at the $K$ point as $\theta$ for crystals with size $r \approx \SI{93}{\micro\meter}$.
    }
    \label{f:bulk_band_simulation}
\end{center}
\end{figure}

Analytically solving the elastic bulk states with re-distributed stress would be difficult. Here we show the simulated bulks modes of the topological insulators (Fig.~\ref{f:bulk_band_simulation}). For a smaller inner triangular hole (smaller $r$) inside the unit cell, we observe a gradually closed and re-opened bandgap when the orientation $\theta$ of the triangle from $-30^{\degree}$ to  $30^{\degree}$, with a topological phase transition at $\theta = 0^{\degree}$ (Figs.~\ref{f:bulk_band_simulation}a,b). For a larger triangular hole size, the modes for smaller orientation angle $\theta$ are distorted, however the topological modes at larger $\abs{\theta}$ are well preserved with protected valley vortex (Fig.~\ref{f:bulk_band_simulation}c). The band and also mode vortices are still inverted for structures with $\theta$ and $-\theta$. This proves that the redistribution of the stress due to our pattern will not break the symmetry of system.    

\clearpage

\section{Dissipation dilution}

\subsection{Numerical modelling}

For the stressed, high-aspect-ratio membranes under consideration, the loss of out-of-plane modes is dominated by bending.
Taking the membrane to lie in the $x-y$-plane, the energy $\Delta W$ dissipated per cycle can be written as\cite{yu2012control,Schmid2023}
\begin{align}
    \Delta W &\approx \frac{2\pi}{Q_\mathrm{int}} W_\mathrm{bend}=
    \\
    &=\frac{\pi D_\mathrm{P}}{Q_\mathrm{int}} \int \left( \partial_{xx} u_z + \partial_{yy} u_z\right)^2 - 2(1-\nu) \left( \partial_{xx}u_z\partial_{yy}u_z - \left(\partial_{xy}u_z\right)^2 \right) \mathrm{d}A,
    \label{eq: energy loss}
\end{align}
where $u(x,y,t) = u_z(x,y) e^{i\Omega_\mathrm{n} t}$ is the out-of-plane displacement of the mode of interest (we have dropped the mode index $n$ for notational simplicity here and in the following).
Furthermore, 
$D_\mathrm{P}= E_1 h^3/(12(1-\nu^2))$ is the flexural rigidity, where $h$ is thickness, $E_1$ and $\nu$ denote the real part of Young's modulus and Poisson's ratio, respectively.

The intrinsic quality factor $Q_\mathrm{int} \equiv {E_1}/{E_2}$ is the ratio of real and imaginary parts of the Young's modulus $E = E_1 + i E_2$ of the resonator material.
The microscopic origins of the dissipation in thin SiN films are not understood in detail.
Yet there is overwhelming evidence that surface effects play a significant role. 
A heuristic model with a volume and a thickness-dependent surface term
\begin{equation}
    Q_\mathrm{int}^{-1}=Q_\mathrm{vol}^{-1}+(\beta h)^{-1}
    \label{e:qint}
\end{equation}
has been shown to match with a wide range of measurements.\cite{Villanueva2014}
We adopt this model and obtain for our films with thickness $h \approx 20~\mathrm{nm} $ a value of $Q_\mathrm{int} \approx 1140$\cite{Villanueva2014}, which we use throughout in our simulations of expected quality factors. 

The stored energy $W$ is dominated by the elongation of the material against the pre-stress in the film.
From the simulated mode profile, it is efficiently computed as the maximal kinetic energy $T$,
\begin{equation}
    W = \max{(T)} = \frac{\rho h \Omega_\mathrm{n}^2}{2} \int u_z^2 \mathrm{d}A,
    \label{eq: stored energy}
\end{equation}
where $\rho$ is the mass density.

The quality factor can then be found as
\begin{equation}
    Q = 2\pi \frac{W}{\Delta W}=D_Q\cdot Q_\mathrm{int},
\end{equation}
where the dissipation dilution factor $D_Q$ is obtained from the mode profile $u_z(x,y)$ in the numerical simulations, and $Q_\mathrm{int}$ is the numerical value obtained from the heuristic model of Eq.~(\ref{e:qint}).

\subsection{Implementation in COMSOL}

We used COMSOL Multiphysics to simulate the phononic topological insulator. 
The following major material parameters were used for the SiN thin film:

\begin{table}[h]
\centering
\begin{tabular}{|l|l|l|}
\hline
Density                         & $\rho $               & $3200~\mathrm{kg/m^3}$       \\ 
Young's modulus                 & $E$                   &   $270~\mathrm{GPa}$     \\ 
Poisson ratio                   & $\nu$                 & 0{.}27     \\ 
In-plane initial stress         & $\bar \sigma$         & $1.145~\mathrm{GPa}$      \\ 
\hline
\end{tabular}
\end{table}
To obtain the dispersion diagram comprising the phononic bulk and edge states, only one unit cell along the propagation direction was considered, using Floquet boundary conditions.
The group velocity of the topological edge modes was extracted from the simulated dispersion with $v_g = {\partial \omega}/{\partial k}$. The dispersion of topological edge modes inside the bandgap is quasi-linear, and the bandgap frequency can differ by 15 kHz from device to device in experiments, therefore we choose a group velocity near the center point of bandgap if not specifically mentioned, that is $v_g = 280 ~\mathrm{m/s}$. In experiments, the different measured bandgap frequencies from different devices indicate that the in-plane stress differs slightly among them, ranging from 1.145 to 1.27 GPa. For simplest, we choose 1.145 GPa for simulation and calculation.

\begin{figure}[ht]
\begin{center}
    \includegraphics[width=\linewidth]{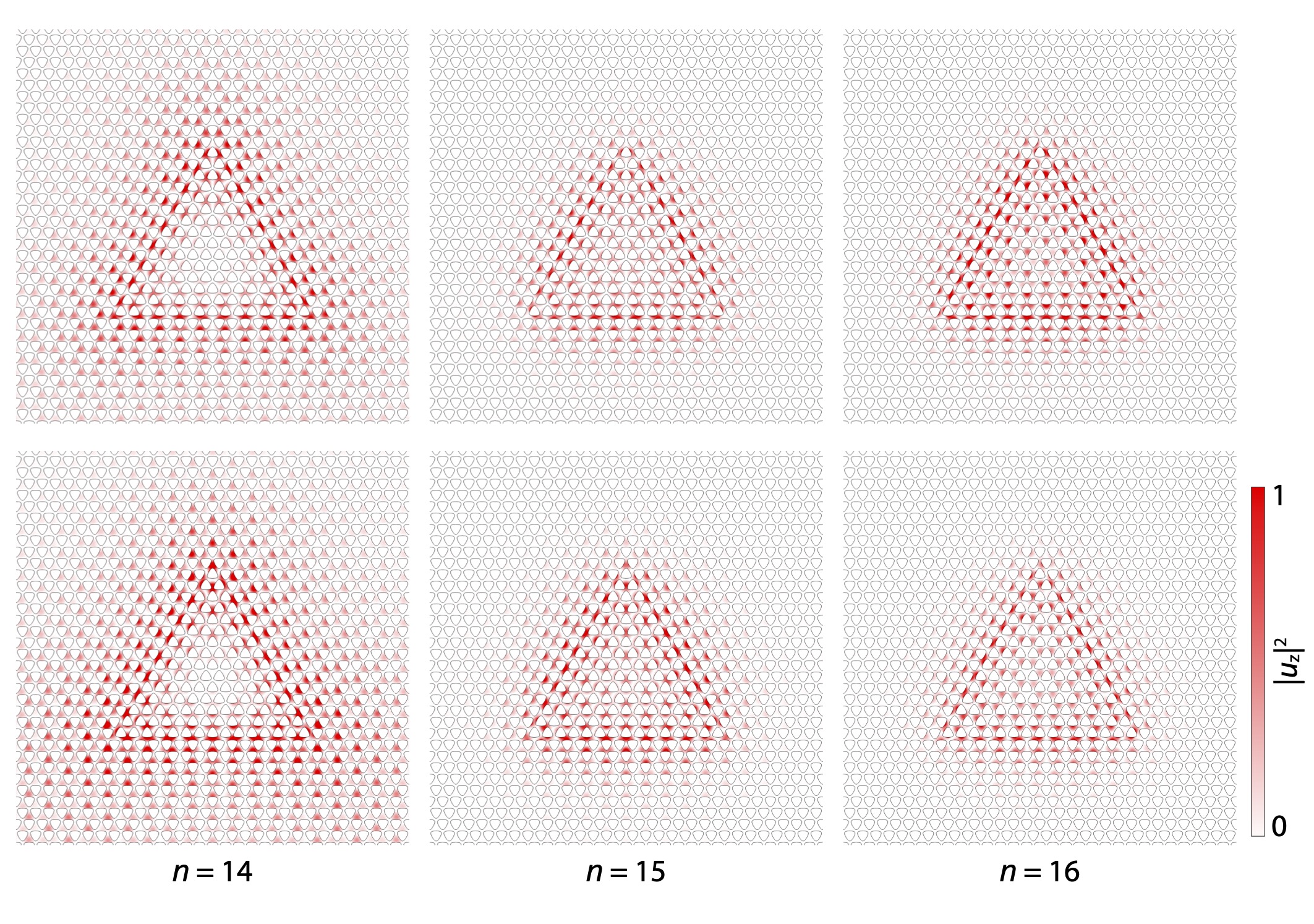}
    \caption{{\bf Simulated modal profiles.} Displacement modal profile of modes with mode number $n=14$, $n=15$ and $n=16$, respectively. The triangle cavity side length is 15 lattice constants. 
    }
    \label{f:simulated_modal_profile}
\end{center}
\end{figure}

To numerically predict losses, we simulated the standing wave resonances in the full triangular phonon cavity with edge length of 15 unit cell, as shown in {Fig. 3a}, whose sides consist of the topological waveguides. 
That is, potential effects of the three corners are taken into account by the simulation.
Example results of the simulation are shown in Fig.~\ref{f:simulated_modal_profile}.

From the solutions of each resonance, we extract the dissipated energy (Eq. \ref{eq: energy loss}) and stored energy (Eq. \ref{eq: stored energy}).
This yields the results listed in Table~\ref{t:simulationResults}.

\begin{table}[htb]
\begin{center}
\begin{tabular}{|l|l|l|l|l|}
\hline
 \begin{tabular}[c]{@{}l@{}}Mode  \\ number $n$ \end{tabular} 
& \begin{tabular}[c]{@{}l@{}}Simulated  \\ Q-factors ($10^6$)\end{tabular} 
& \begin{tabular}[c]{@{}l@{}}Best measured\\ Q-factors ($10^6$)\end{tabular} 
& \begin{tabular}[c]{@{}l@{}}Simulated\\ splitting (Hz)\end{tabular} 
& \begin{tabular}[c]{@{}l@{}}Smallest measured  \\ splitting (Hz)\end{tabular}
\\ 
\hline
14          & 22.9/25.6                                                    & 7.4/7.0                                                         & 424                   & 232                                                                      \\ \hline
15          & 11.1/20.1                                                     & 39.0/25.2                                                       & 209                   & 277                                                                      \\ \hline
16          & 9.8/10.2                                                      & 27.3/22.4                                                       & 3                     & 17                                                                       \\ \hline
17          &                                                               & 14.8/8.6                                                        &                       &                                                                          \\ \hline
\end{tabular}
\caption{Simluated and measured quality factors and mode splittings for the full triangular cavity. The two numbers of Q correspond to the split mode pairs}
\label{t:simulationResults}
\end{center}
\end{table}

It should be noted that the $n=14$ and $n=17$ modes are spectrally located at the band edge. 
Especially for the $n=17$ mode the simulations showed hybridization with modes in the bulk.
Therefore, their parameters were not evaluated numerically.

\begin{figure}%[ht]
\begin{center}
    \includegraphics[width=\linewidth]{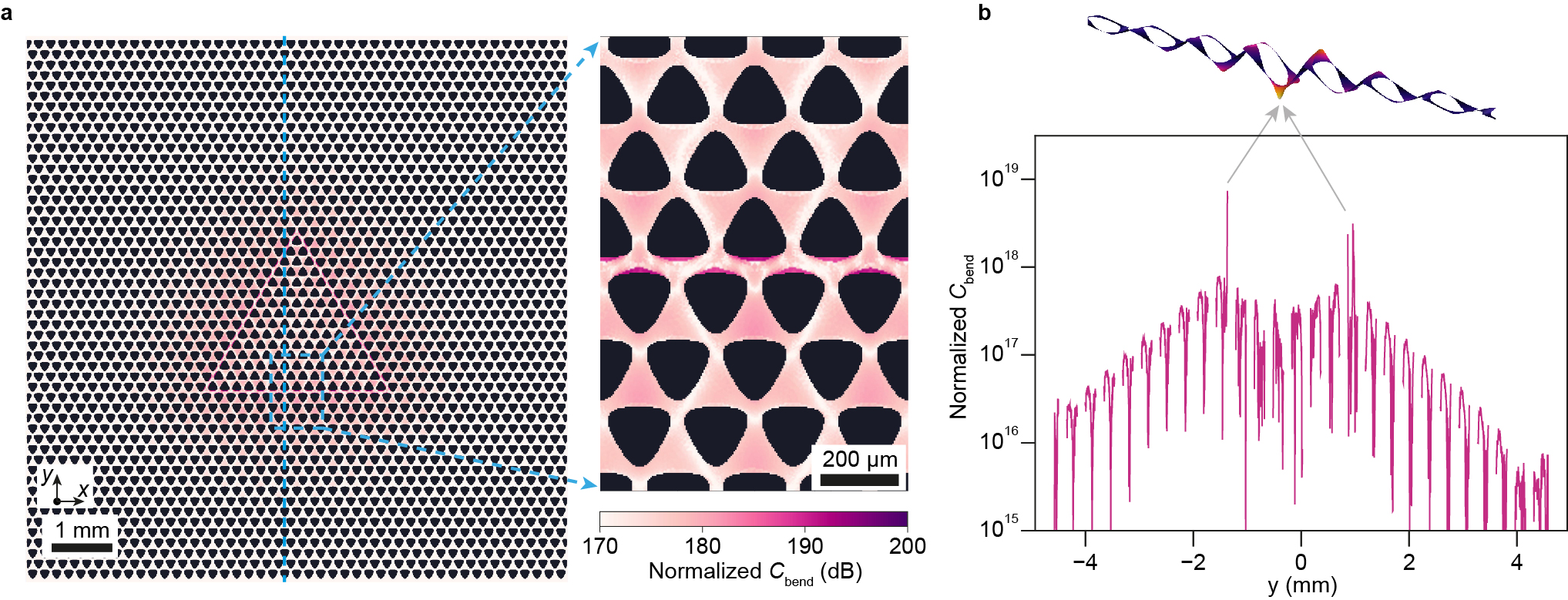}
    \caption{{\bf Simulated bending curvature.} (a) Simulated bending curvature $C_\mathrm{bend}$ of the whole device (left) and part of the device (right). (b) Simulated bending curvature along the vertical dashed blue line in (a). The top inset shows the simulated topological edge modes near the domain interface. The bending curvature is normalized to the maximal kinetic energy $W$ in Eq. (\ref{eq: stored energy}).
    }
    \label{f:simulaed_curvature}
\end{center}
\end{figure}

As discussed above, the loss is dominated by the bending of the mode. 
To estimate the source of the loss, the bending curvature distribution of one of the modes ($n = 15$) is also calculated, with the bending curvature $C_\mathrm{bend} = \left( \partial_{xx} u_z + \partial_{yy} u_z\right)^2 - 2(1-\nu) \left( \partial_{xx}u_z\partial_{yy}u_z - \left(\partial_{xy}u_z\right)^2 \right)$ from Eq. (\ref{eq: energy loss}). 
Figure \ref{f:simulaed_curvature}a shows that the dominating bending curvature is localised near the topological domain interface.
The curvature near the membrane's rim to the substrate is orders of magnitude smaller than that near topological domain interface. 
The large curvature at the exact domain interface is much larger than that at any other position (peaks in Fig. \ref{f:simulaed_curvature}b).
This comes from the extra bending of the topological edge states (inset of Fig. \ref{f:simulaed_curvature}b).
Integrating the peak over the entire membrane, however, reveals that the bending at the domain interface contributes only around 20\% of the total loss.
For the other modes in Fig. \ref{f:simulated_modal_profile}, the contribution ranges from 10\% to 40\%.
With a proper design of the edge, for example, gradually increasing the triangular size $r$ from the interface, the extra bending loss may be reduced.

\subsection{Comparison with hard-clamped waveguide}
\begin{figure}[hp]
\begin{center}
    \includegraphics[width=\linewidth]{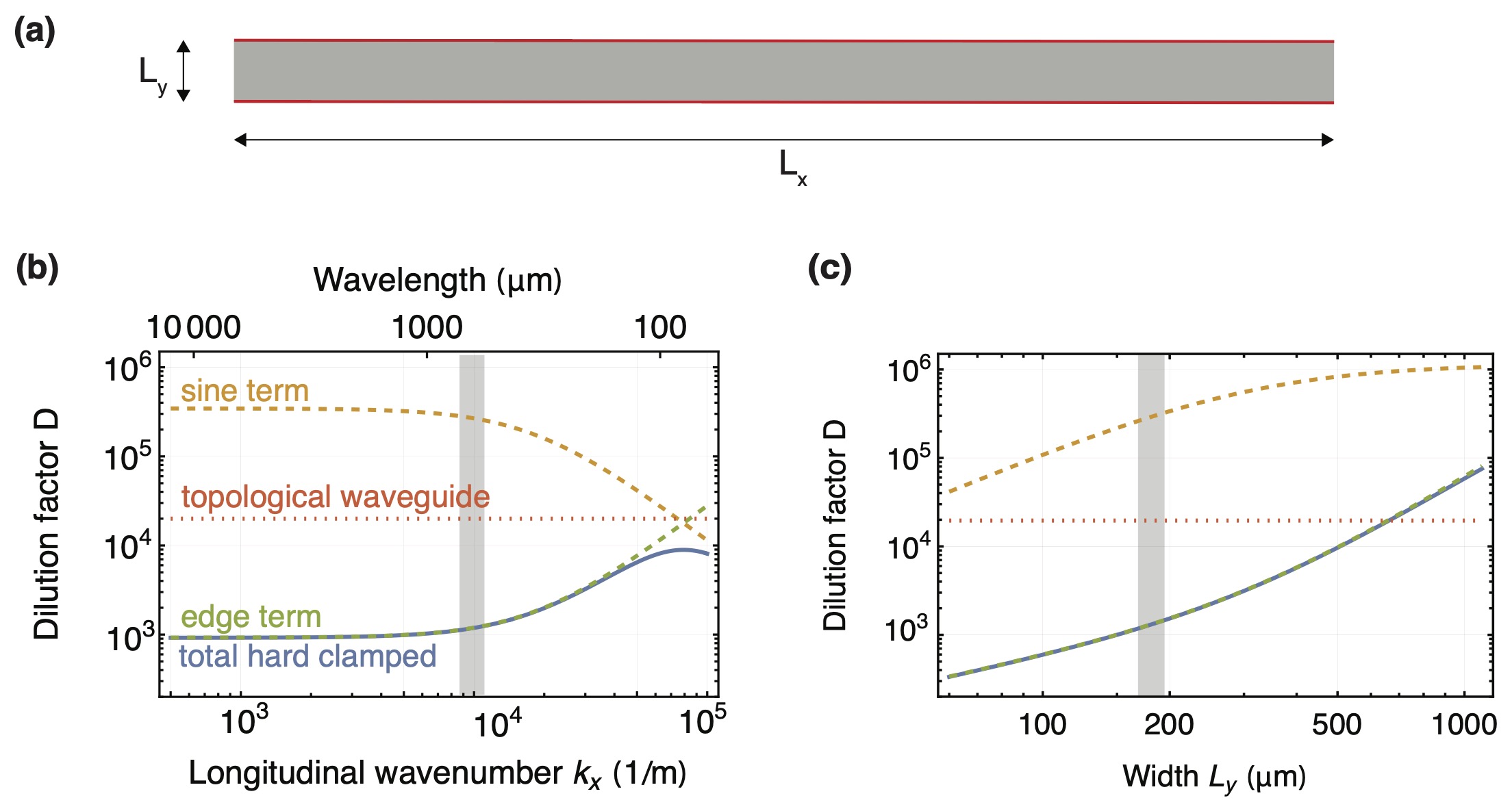}
    \caption{{\bf Dissipation dilution in a hard-clamped waveguide.} 
    (a) Geometry of the waveguide with width $L_y$ and length $L_x$, subject to hard clamping along  $L_x$, as indicated with a red line.
    (b) Dilution factor $D_Q$ for fixed width $L_y=\SI{170}{\micro\meter}$, with contributions from the sine and edge terms, as well as comparison with the simulated dilution factor of the topological waveguide.
    Gray bar indicates approximate longitudinal wavenumber of the topological waveguide in the $n=15$ mode of the triangular cavity.
    (c) Dilution factor $D$ for fixed longitudinal wave number $k_x=10^4~/\mathrm{m}$. Same color code as in (b).
    Gray bar indicates widths corresponding to an approximate transverse wavenumber $L_y=\pi/k_\perp$ of the topological waveguide.
    }
    \label{f:hardClamp}
\end{center}
\end{figure}

To benchmark the topological phonon waveguides, we compare their attained loss with that of a standard ``hard-clamped'' waveguide as described in ref.~\cite{Romero2019}.
To model standing-wave modes on such a waveguide, we start from the analytical expression\cite{Schmid2023} for the dissipation dilution factor 
\begin{equation}\label{eq:hcdil}
    D_Q= \left[
       \underbrace{\frac{D_\mathrm{P}}{h \Bar{\sigma}}\left( k_x^2 + k_y^2\right)}_\text{sine term}
    + \underbrace{2\sqrt{\frac{D_\mathrm{P}}{h \Bar{\sigma}}}\frac{L_x k_y^2 + L_y k_x^2}{L_x L_y \left(k_x^2 + k_y^2\right)}}_\text{edge term}
    \right]^{-1}
\end{equation}
of a rectangular, stressed membrane of sidelengths $L_x$ and $L_y$ (see Fig.~\ref{f:hardClamp}), where $k_{\{x,y\}} = \pi n_{\{x,y\}}/L_{\{x,y\}}$ is the wavenumber and $n_{\{x,y\}}$ the mode index in the ${\{x,y\}}$ direction respectively.

In Eq.~(\ref{eq:hcdil}), the first term in the bracket is due to the sinusoidal bending of the membrane in its bulk, whereas the second term is related to the sharp kink that forms at the ``hard clamp'' of the membrane to its substrate at the membrane edge.\cite{Schmid2023}
In the case of the waveguide considered here, the structure is translationally symmetric in the $x$-direction, and there is clamp along the vertical direction as shown in Fig.~\ref{f:hardClamp}a.
This prompts us to eliminate one of the edge terms, keeping only the contribution of the membrane edge along the $x$-direction (see Fig.~\ref{f:hardClamp}a), yielding
\begin{equation}
    D_Q= \left[
       {\frac{D_\mathrm{P}}{h \Bar{\sigma}}\left( k_x^2 + k_y^2\right)}
    + {2\sqrt{\frac{D_\mathrm{P}}{h \Bar{\sigma}}}\frac{ k_y^2}{ L_y \left(k_x^2 + k_y^2\right)}}
    \right]^{-1},
\end{equation}
which does not depend on $L_x$ any more, but only on the waveguide width $L_y$ and the longitudinal ($k_x$) and transverse ($k_y$) wavenumbers.
For the fundamental transverse mode we have $L_y=\pi/k_y$ and eventually 
\begin{equation}\label{eq:hcdilwg}
    D_Q= \left[
       {\frac{D_\mathrm{P}}{h \Bar{\sigma}}\left( k_x^2 + k_y^2\right)}
    + {\frac{2}{\pi}\sqrt{\frac{D_\mathrm{P}}{h \Bar{\sigma}}}\frac{ k_y^3}{ k_x^2 + k_y^2}}
    \right]^{-1}.
\end{equation}

From the simulated mode pattern of the $n=15$ modes in the topological waveguide, we extract longitudinal and transverse wavenumbers of  $k_\parallel\approx 1{.}0\cdot 10^4/\mathrm{m}$ and $k_\perp \approx 1{.}8\cdot 10^4/\mathrm{m}$.
Since these wavenumbers encode zero-crossings as well as membrane curvature, we evaluate the dissipation dilution factor of a hard-clamped membrane with $k_x=k_\parallel$ and $k_y=k_\perp$ for comparison,
yielding
\begin{equation}\label{eq:hcdilwg}
    D_Q\approx \left[
        \frac{1}{265000}+\frac{1}{1200}
    \right]^{-1}\approx 1195.
\end{equation}

With a dissipation dilution factor of $D_Q\approx 2\cdot 10^4$ for the topological waveguide mode, this implies that it performs significanly better than a comparable hard-clamped waveguide, whose dissipation is dominated by the edge term.
It is also clear, however, that the soft-clamping of the topological waveguide is not ideal, in the sense that its dilution factor is still well below what can be achieved in a situation where there is only the sine term.

In Fig.~\ref{f:hardClamp} we plot the hard-clamped waveguide's dilution factor for a broader range of parameters. 
As expected, the dissipation dilution of the hard-clamped rectangular waveguide is limited by the edge term for most of the parameter range of interest (except for very large longitudinal wavenumbers).
Furthermore, the topological waveguide shows better dilution than a hard-clamped device over a wide parameter range.
Nonetheless there is margin for improvements with respect to the pure soft-clamping limit represented by the sine term.

In such a comparison it should be noted that hard-clamped devices in many cases do not reach $Q$-factors limited by the diluted internal dissipation as discussed above. 
For example, for the hard-clamped, rectangular waveguide of width $L_y=\SI{77}{\micro\meter}$, length $L_x=1.8~\mathrm{cm}$ and thickness $h=80~\mathrm{nm}$ considered in ref.~\cite{Romero2019} the dilution factor (\ref{eq:hcdil}) evaluates to $D_Q\approx 160$ for a mode with $k_x=3.4\cdot10^4/\mathrm{m}$.
Together with an estimated $Q_\mathrm{int}\approx 4100$ for this thickness, the expected quality factor is $Q=6\cdot 10^5$.
However, the measured quality factor is only $\sim 2000$.
The large discrepancy is likely due to radiation loss, to which hard-clamped devices are prone---in contrast to the topological edge waveguide protected by a bandgap.

\section{Backscattering-induced mode splitting}
\subsection{Mode splitting caused by a single scatterer}
\label{ss:single}

\begin{figure}[htb]
\begin{center}
    \includegraphics[width=0.6\linewidth]{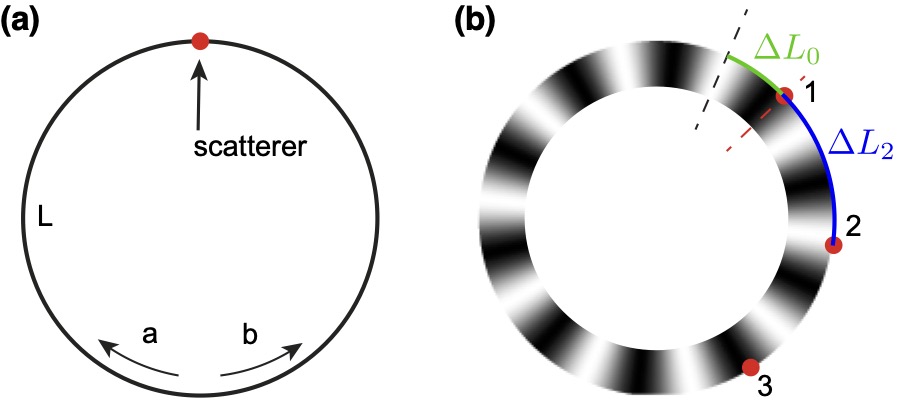}
    \caption{{\bf Backscattering in a travelling-wave resonator.} (a) Single scatterer. Here, $a$ and $b$ are the clockwise and counter-clockwise propagating modes, which get coupled by the scatterer. The cavity length is $L$.
    (b) Multiple scatterers 1-3, located at different positions with respect to the standing wave pattern of one of the supermode that forms.}
    \label{f:backscattering}
\end{center}
\end{figure}

We first consider a simple case when there is one scatterer inside a traveling wave resonator of round-trip length of $L$, as shown in Fig.~\ref{f:backscattering}.
The scatterer has transmission and reflection coefficients $t$ and $r$, with $tt^* + rr^* = 1$. 
In the absence of backscattering, there exist two degenerate counter-propagating resonant modes $a$ and $b$. 
Backscattering couples the two modes, thereby breaking the degeneracy and forming a pair of supermodes. 

To calculate the induced mode splitting, we use a transfer matrix model similar to the one discussed in ref.~\cite{Ren2022topo}.
The loop matrix for one round-trip in the cavity is given by 
\begin{align}
M=T_\mathrm{scatt} T_\mathrm{edge}
=\begin{pmatrix}
    1/t^* & -r^*/t^*\\
    -r/t    & 1/t
\end{pmatrix}
\begin{pmatrix}
    e^{-i k L} & 0\\
    0    & e^{+k L}
\end{pmatrix}.
\end{align}
The periodic boundary condition $\det{(M-\mathbb{1})}=0$ translates to
\begin{equation}
  \cos(k L)=\sqrt{1-r^2}
\end{equation}
for real $r$ and $t$, which we assume for simplicity.
In the absence of backscattering ($r=0$), the allowed wavenumbers are evidently given by $k_n = 2\pi n/L$, with $n$ integer.
Perturbations by a small $r\ll1$ will lead to a  shift $k_n\rightarrow k_n+\delta k_n$ satisfying
\begin{equation}
  \cos( (k_n+\delta k_n) L)=\sqrt{1-r^2}.
\end{equation}
Taylor-expanding both sides to second order yields the solutions
\begin{equation}
  \delta k_n^{\pm}=\pm \frac{r}{L},
\end{equation}
which implies a splitting of $\Delta k_n=\delta k_n^{+}-\delta k_n^{-}=2r/L$.
Finally, we use the group velocity $v_\mathrm{g}=d\Omega/dk$, to  obtain the frequency splitting
\begin{equation}
  \Delta \Omega_n=v_\mathrm{g} \Delta k_n=\frac{2 v_\mathrm{g} r}{L}.
\end{equation}
Conversely, the backscattering probability $p_\mathrm{bs}$ can be obtained from the measured splitting (using also the definition of the free spectral range $\Omega_\mathrm{FSR}/2\pi= v_\mathrm{g}/L$) as 
\begin{equation}
  p_\mathrm{bs} = |r|^2 = \left(\frac{\Delta\Omega_n L}{2v_\mathrm{g}}\right)^2=\left(\pi\frac{\Delta\Omega_n }{\Omega_\mathrm{FSR}}\right)^2.
\label{eq:backscatteringProbability}
\end{equation}

\subsection{Mode splitting caused by multiple scatterers}
\label{ss:multiple}
In the presence of $N$ sources of (weak) backscattering, the backscattered waves interfere.
The net result is still the hybridization of counterpropagating waves to standing-wave supermodes, with their degeneracy lifted.
For weak scattering, this scenario has been analysed in detail in the context of optical whispering gallery mode resonators\cite{Lina2013njp}, and we adapt the results here.

Dropping the longitudinal mode index $n$ for notational convenience, the total splitting
\begin{equation}
  \Delta \Omega \equiv \left | \delta \Omega^+ -\delta \Omega^-\right |
\end{equation}
can be obtained from the frequency shift of the two supermodes
\begin{align}
    \delta \Omega^+ &= \sum_{i=1}^N \Delta\Omega^{(i)} \cos^2(\psi_0-\phi_i)\\
    \delta \Omega^- &= \sum_{i=1}^N \Delta\Omega^{(i)}\sin^2(\psi_0-\phi_i).
\end{align}
Here, the index $i$ enumerates the $N$ scatterers and $\Delta\Omega^{(i)}$ is the splitting that would be observed if only the $i$-th scatterer was present.
Furthermore, $\psi _0$ denotes the spatial phase distance ($\sim k \Delta L_0$) between the first scatterer and the anti-node of one of the supermodes, and $\phi _i$ is the spatial phase distance ($\sim k \Delta L_i$) between the $i$th scatter and first scatter, see Fig.~\ref{f:backscattering}.
By definition, $\phi_1=0$.
The spatial phase $\psi _0$ itself depends on the distribution and properties of the scatterers, and will adjust so as to maximize the spliting.\cite{Lina2013njp}

Assuming that the backscattering in our phononic cavities is dominated by the three corners, each scattering back with approximately equal amplitude, we obtain
\begin{align}
  \Delta\Omega^{(\angle)}&\equiv\Delta\Omega^{(1)}=\Delta\Omega^{(2)}=\Delta\Omega^{(3)}\\
  \phi_1&=0\\
  \phi_2&= (1/3)\cdot 2 \pi n\\
  \phi_3&= (2/3)\cdot 2 \pi n ,
\end{align}
yielding 
\begin{equation}
\label{e:finalBS}
    \Delta\Omega_n=\begin{cases}
      3\Delta \Omega^{(\angle)}, & \text{if}\ (n~\mod 3)=0 \\
      0, & \text{otherwise}.
    \end{cases}
\end{equation}
Here, for the case of a mode number $n$ divisible by 3, we have already maximized the splitting over $\psi_0$.

The form of Eq.~(\ref{e:finalBS}) is straightforwardly explained with the physics of interference.
For a mode number divisible by three, the backward scattered waves interfere constructively, and their amplitudes ($\propto \Delta \Omega^{(\angle)}$) add up, whereas in the other cases the waves interfere destructively.

 In our experiment, $\Delta \Omega_{n=16}\neq 0$.
 The likely reason is that the $\Delta\Omega^{(i)}$ are not exactly equal for the three corners; the phases $\phi _i$ might also differ slightly.
 However, averaging over several samples, we observe that 
 $\langle\Delta \Omega_{n=16}\rangle$ is more than 4 times smaller than the average splitting $\langle\Delta \Omega_{n=15}\rangle$ (see Figure 4 in the main manuscript).
 The splittings of the  $n = 14$ and $n=17$ modes are only slightly lower than that of $n= 15$
 (see Table~\ref{t:simulationResults}).
These are close to the bandgap edge and tend to hybridize with the bulk modes of the devices, which can explain their larger splitting.

For the triangular cavity with 15 unit cells per side, the total round-trip length is $L\approx\SI{9300}{\micro\meter}$, and the mid-gap group velocity is $v_\mathrm{g} = 280$ m/s.
The average back-scattering-induced splitting per corner is $\langle \Delta\Omega_{n=15}\rangle/3\approx 2\pi\cdot 102 \,\mathrm{Hz}$, yielding a backscattering probability in one corner of  $p_\mathrm{bs}^{(\angle)} = 1{.}1\cdot 10^{-4}$ per eq.~(\ref{eq:backscatteringProbability}). 

\subsection{Comparison with a trivial hard-clamped waveguide cavity}
\begin{figure}[htb]
\begin{center}
    \includegraphics[width=0.8\linewidth]{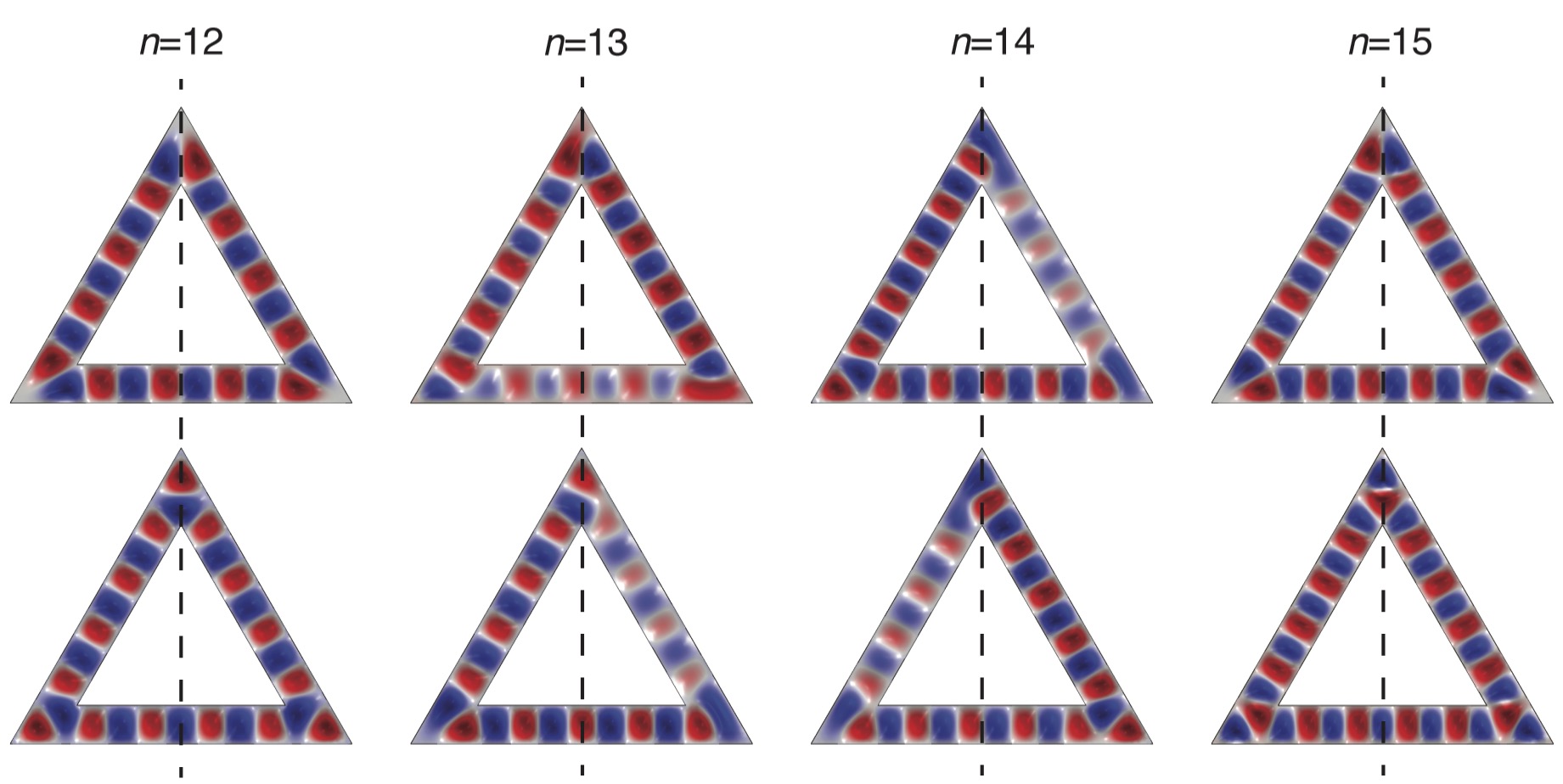}
    \caption{{\bf Simulated mode profile of a trivial triangle phonon cavity.} The thickness and material properties are the same as the membranes used in this paper.
    }
    \label{f:trivial_tri_backscattering}
\end{center}
\end{figure}

For comparison, we again analyse the case of the hard-clamped ribbon waveguide. 
We perform a simulation of the eigenmodes of a triangular cavity, in which such a waveguide is wrapped up.
The ribbon width is $\SI{420}{\micro\meter}$, and the length of the triangle cavity is approximately the same as the topological triangle cavities with edge length of 15 unit cells.
Thereby, the obtained modes have similar wavelength and  frequency as the toplogical cavities with the same mode order.
The results are shown in Fig.~\ref{f:trivial_tri_backscattering} and table~\ref{tab:trivial triangle cavity}.

\begin{table}[h]
    \centering
\begin{tabular}{|l|r|r|r|r|}
\hline
mode number $n$ & 12      & 13      & 14      & 15      \\
\hline
$\Omega^+/2\pi$ (Hz)   & 1{,}174{,}765 & 1{,}189{,}675 & 1{,}254{,}329 & 1{,}326{,}117 \\
\hline
$\Omega^-/2\pi$(Hz)      & 1{,}120{,}234 & 1{,}189{,}624 & 1{,}254{,}300 & 1{,}269{,}252 \\
\hline
$(\Omega^+-\Omega^-)/2\pi$ (Hz)   & 54{,}531   & 51      & 29      & 56{,}865  \\
\hline
\end{tabular}
    \caption{Simulation of the frequency and mode splitting of a trivial triangle cavity, which is corresponding to Fig. \ref{f:trivial_tri_backscattering}}
    \label{tab:trivial triangle cavity}
\end{table}

The qualitative behaviour is comparable, with mode pairs $\Omega_n^\pm$, whose splitting is larger for $n \mod 3=0$ and smaller otherwise.
Yet the quantitative scale is entirely different, with the $n=12$ and $n=15$ modes featuring splittings approaching the free spectral range.

For the large backscattering underlying this splitting, the perturbative model of section~\ref{ss:multiple} breaks down.
However, we can adapt the transfer matrix model of section~\ref{ss:single}, using\cite{Ren2022topo}
\begin{align}
  M &=T_\mathrm{scatt} T_\mathrm{edge}T_\mathrm{scatt} T_\mathrm{edge}T_\mathrm{scatt} T_\mathrm{edge}
\intertext{and}
  T_\mathrm{edge} &=
\begin{pmatrix}
    e^{-i k L/3} & 0\\
    0    & e^{+k L/3}
\end{pmatrix}.
\end{align}
Again for real $r$ and $t$ the periodic boundary condition reads
\begin{equation}
    \left(t-\cos\left(\frac{k L}{3} \right) \right)\left(t+2 \cos\left(\frac{k L}{3}\right)\right)^2=0.
\end{equation}
Solving this equation graphically for allowed wavenumbers $k$, and comparing it with the simulated frequencies (assuming constant group velocity $v_\mathrm{g}$) suggests a backscattering probability $p_\mathrm{bs}\sim 0{.}8$ at each corner.

\section{Parametric amplification}

\begin{figure}[h]
\begin{center}
\includegraphics[width=0.5\linewidth]{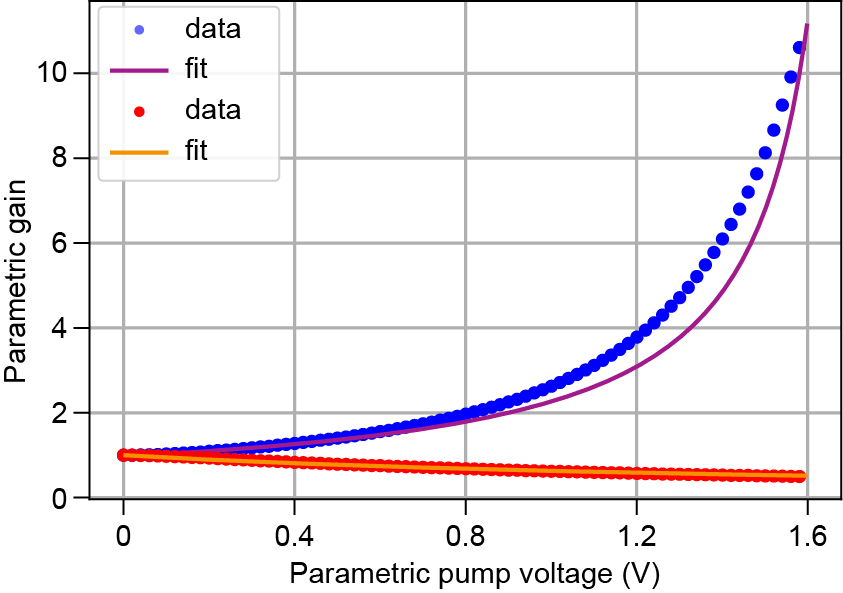}
\caption{{\bf Parametric amplification and damping}. Measured parametric gain G for in-phase $\psi = 0^{\degree}$ and out-of-phase $\psi = 90^{\degree}$ parametric drive.}
\label{f:subfig_parametric} 
\end{center}
\end{figure}

Thin-membrane-based mechanical devices feature with strong nonlinearity and has been explored for applications like signal amplification or noise squeezing. Our topological devices inherit this merit. Based on that, we demonstrated phase-sensitive topological amplification, as shown in Fig. \ref{f:subfig_parametric}. Generally, parametric amplification or squeezing of a resonant mechanical mode of frequency $\Omega_n$ can be realized by adding an external force with frequency of $2\Omega_n$ to the system \cite{Rugar1991, Xiang2024}. In our experiments, this was realized by adding a parametric pump voltage $V_{2\Omega_n}\cos{(2\Omega_nt)}$ to the piezoelectric actuator that exerts force to the membranes (Fig. ~\ref{f:subFig_setup}). A small mechanical signal to be amplified was realized by adding a small voltage $V_\mathrm{signal}\cos{(\Omega_nt + \psi)}$ to the piezoelectric actuator. Here $\Omega_n$ was chosen to be the frequency of one of the topological edge modes. The displacement gain of this small signal provided by the parametric pump is\cite{Rugar1991}:
\begin{equation}
G=\frac{|x_\mathrm{on}|}{|x_\mathrm{off}|}=\left[\frac{\cos^2{\psi}}
{(1+V_{2\Omega_n}/V_{th})^2}+\frac{\sin^2{\psi}}
{(1-V_{2\Omega_n}/V_{th})^2}\right]^{\frac{1}{2}},
\label{Eq. para_gain}
\end{equation}
where $|x_{on}|$, $|x_{off}|$ are absolute values of mechanical displacement with and without parametric force, respectively. $V_{th}$ is the threshold voltage for reaching the region of instability for parametric oscillation. This threshold is proportional to the mechanical dissipation rate and inverse proportional to the nonlinear coefficient \cite{Xiang2024, Patil2015}. As the Q-factors of topological edge modes in our devices are able to reach $38\cdot10^6$, we have a potential to reach very low threshold at room temperature. 

In this preliminary demonstration, we choose a topological edge mode  with Q around $1.8 \cdot 10^7$. 
A systematical study of the parametric amplification and oscillation on high-Q topological egde modes will be presented in our future study. 
Figure  \ref{f:subfig_parametric} shows the results of the measurements with the analytical fits. From the fits we extract $V_\mathrm{th}\approx1.7~\mathrm{V}$. A further reduction of this threshold can realised by gluing the chip to the piezoelectric actuator or using electromechanical configurations. \cite{Xiang2024,Mashaal2024}.

\section{Additional data}

\subsection{Corner scans}
\begin{figure}[h]
\begin{center}
\includegraphics[width=\linewidth]{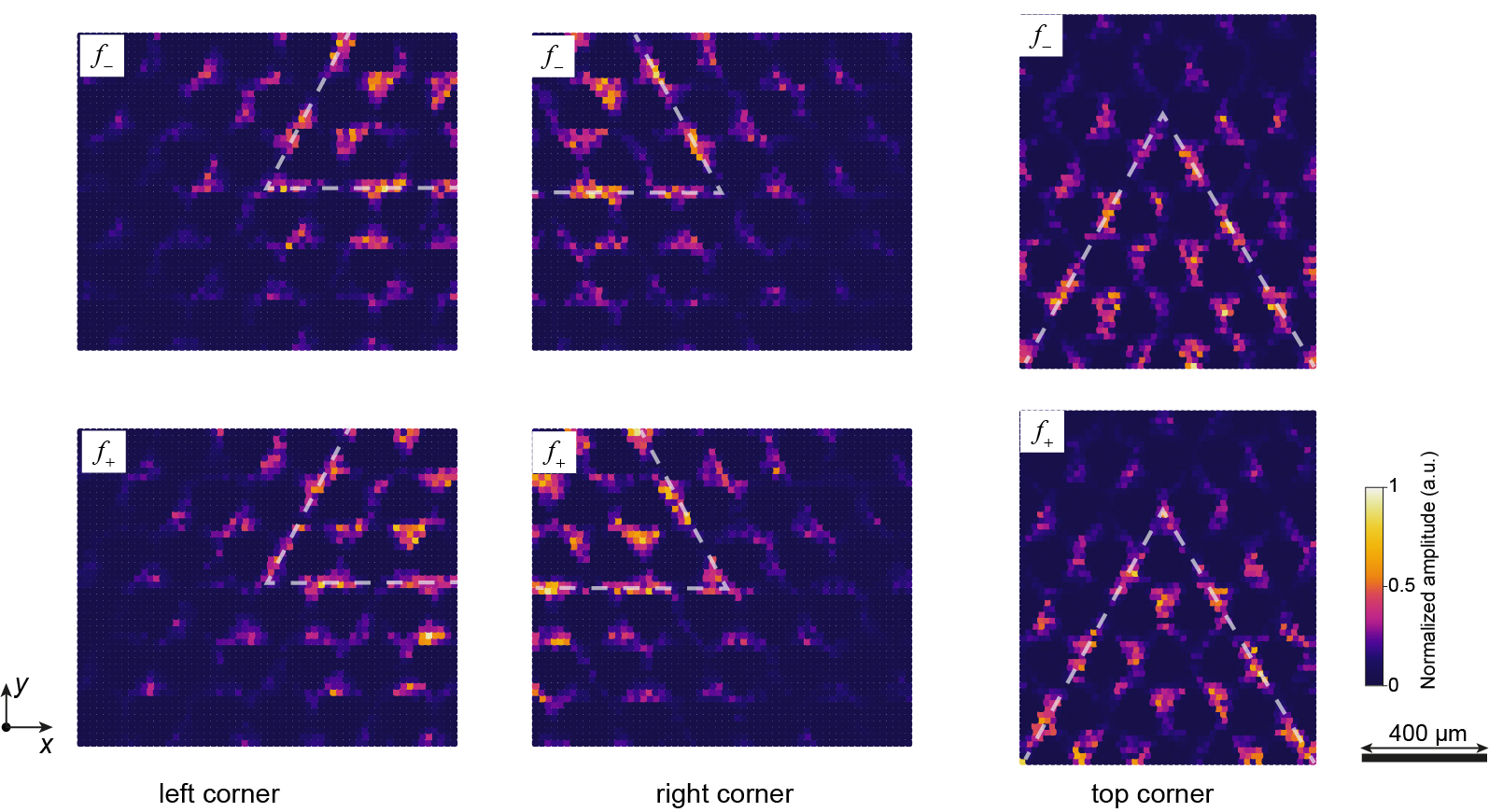}
\caption{{\bf Measured spatial amplitude distribution of a pair of split modes with mode number 15 around corners.} The device is one of the devices from {Fig.4d}. $f_-$ and $f_+$ represent the mode with lower frequency and higher frequency, respectively. The left (middle, and right) two panels are the results of scanning across the left (right, and top) corner of the cavity.}
\label{f:corner_scan}
\end{center}
\end{figure}

Figure \ref{f:corner_scan} shows the measured spatial amplitude distribution of a pair of split modes from one of the devices in {Fig. 4d} around the three corners of the triangular cavity. The mode number $n$ is 15. It clearly shows that the corners of the triangle are "locked" to the nodes of the modal profile for one mode while they are "locked" to anti-nodes for another. Besides, one can also see the modes spatially decay fast outside the cavity.

\subsection{Edge scans}
\begin{figure}[h]
\begin{center}
\includegraphics[width=0.7\linewidth]{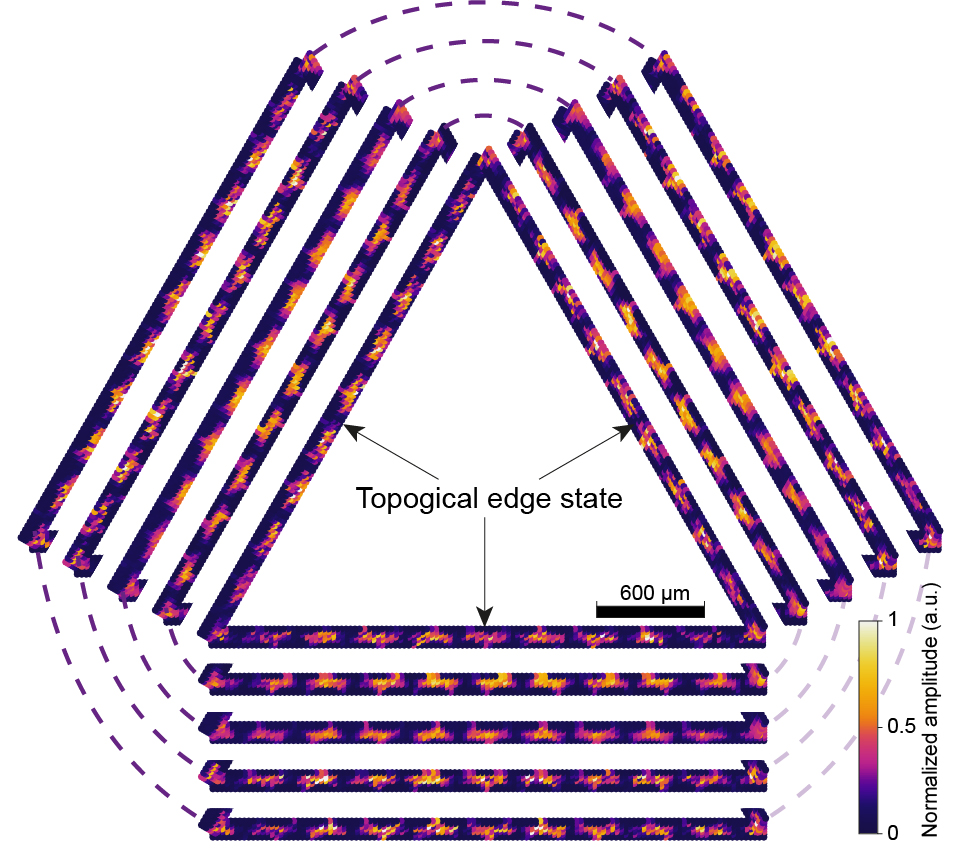}
\caption{{\bf Measured spatial amplitude distribution of one mode of split modes with mode number 15.} The devices are the same as devices from {Fig.4d}. The mode is one of the other split modes corresponding to the one in {Fig.4d}.}
\label{f:tri_edge_scan_antinode}
\end{center}
\end{figure}

Figure \ref{f:tri_edge_scan_antinode} shows the measured spatial amplitude distribution of one mode from a pair of split modes around the edge of the triangular cavity. Their mode number $n$ is 15, and they correspond to the partner modes of the modes in {Fig. 4d}. It clearly shows their modes are spatially orthogonal to that in {Fig. 4d}. The nodes and anti-nodes spatially align with each other, and the three corners are all "locked" to the anti-nodes of the mode. 

\end{document}